# Solar cell efficiency, diode factor and interface recombination: insights from photoluminescence

*Taowen Wang[*], Florian Ehre, Thomas Paul Weiss, Boris Veith-Wolf, Valeriya Titova, Nathalie Valle, Michele Melchiorre, Jan Schmidt, and Susanne Siebentritt*

T. Wang, Dr. F. Ehre, Dr. T. P. Weiss, Dr. M. Melchiorre and Prof. Dr. S. Siebentritt
Laboratory for Photovoltaics (LPV), Department of Physics and Materials Science, University of Luxembourg, 41 rue du Brill, L-4422, Belvaux, Luxembourg.
E-Mail: taowen.wang@uni.lu

Dr. B. Veith-Wolf, Prof. Dr. J. Schmidt and Dr. V. Titova
Institute for Solar Energy Research Hamelin (ISFH), Am Ohrberg 1, D-31860 Emmerthal, Germany.

Prof. Dr. J. Schmidt
Institute of Solid-State Physics, Leibniz University Hannover, Appelstr 2, 30167 Hannover, Germany

Dr. N. Valle
Luxembourg Institute of Science and Technology (LIST), Materials Research and Technology Department, 41 rue du Brill, L-4422, Belvaux, Luxembourg.

**ABSTRACT**

Metastable defects can decisively influence the diode factor and thus the efficiency of a solar cell. The diode factor is also influenced by the doping level and the recombination mechanisms in the solar cell. Here we quantify how the various parameters change the diode factor by photoluminescence measurements and simulations. In addition, we show that backside recombination reduces the open circuit voltage in $CuInSe_2$ solar cells by more than 40 mV. Passivation by a Ga gradient is shown to be as efficient as a passivation by dielectric layers. Increased backside recombination reduces the diode factor, not because of less metastable defect transformation but because of a sublinear increase in photo generated carriers with excitation. This reduction in diode factor is unwanted, since the increased recombination reduces the voltage. A higher doping level, on the other hand, reduces the diode factor, thereby increasing the fill factor, and at the same time increases the voltage.

## 1. Introduction

Many semiconductors that are used as absorbers in solar cells contain defects that show metastable behavior.[1-3] This metastable behavior is often detected as persistent photoconductivity.[2, 4-6] A prominent mechanism is the transformation of donor states into an acceptor states upon electron capture.[1] The back transformation has a considerable activation energy and thus takes time, making the transformation metastable. This transformation leads to an additional shift of the majority carrier Fermi level, leading to an increase in the diode factor, thereby decreasing the fill factor and the efficiency of the solar cell.[7] Besides the presence of metastable defect the various recombination channels in the bulk of the absorber and at the interfaces determine the efficiency. While it is well-known that increased recombination reduces the open circuit voltage, the interplay between increased recombination, e.g. at the back contact and the diode factor is less obvious. In a complete device the factors influencing recombination and diode factor are numerous. It is thus desirable to study the absorber before finishing the solar cell. Photoluminescence offers a tool to study the quasi-Fermi level splitting, i.e. the open circuit voltage, as well as the diode factor.[8] Thus, by photoluminescence it is possible to study the interplay between bulk recombination, back side recombination, diode factor and efficiency, before adding more complexity by introducing the contact layers. Here we concentrate on Copper Indium Selenide (CuInSe$_2$) solar cells, because it has been shown that mitigating back contact recombination is critical to achieve high efficiencies.[9] We study the influence of back side recombination on the quasi-Fermi level splitting ($\Delta E_F$), i.e. the open circuit voltage ($V_{oc}$), as well as on the diode factor. Furthermore, we study the influence of the doping level on the diode factor.

CuInSe$_2$ solar cells suffer a significant open-circuit voltage loss due to a very high back surface recombination velocity of more than $10^5$ cm/s,[10] which can be mitigated by a Ga back grading (GBG).[9] Empa has realized a notable GBG sample with a narrow bandgap around 1.0 eV and efficiency of 16.1% in 2018,[9, 11] an even more outstanding result of 19.6% was achieved in 2019 by increasing the Cu/III ratio and incorporating a Rb post-deposition treatment (PDT).[12] Regarding CIGSe$_2$ solar cells, the Ga gradient is also a classical strategy to improve efficiency and make record solar cells.[13] However, the incorporation of Ga may lead to inhomogeneous lateral distribution of the group-III elements Ga and In, which results in bandgap fluctuations. The potential influence of bandgap fluctuation on photovoltaic performance of solar cells has already been investigated.[14] Bandgap fluctuations lead to tail states and reduce the $V_{oc}$.[15-17] In addition, increase in the bandgap due to incorporation of Ga introduces more abundant donor defects located at 1.1 eV above the valence band edge[18] and makes the well-known defects around 0.8 eV approach deeper into middle gap,[19-20] which finally induces degradation of the minority carrier lifetime as well as the performance of solar cells. To avoid these drawbacks caused by Ga, dielectric metal oxides passivation layers can be applied to reduce back surface recombination. The dielectric metal oxides layers have successfully been applied in c-Si photovoltaics to reduce the surface recombination.[21-22] For CIGSe$_2$ solar cells, a high negative fixed charges $Q_f$ of approximately $10^{13}$ cm$^{-2}$ was experimentally detected between CIGSe$_2$/Al$_2$O$_3$ interface, which lowers the minority carrier density at this interface by roughly 8 orders of magnitude, thus reducing interface recombination [23]. Passivation effects were also demonstrated in devices, because $V_{oc}$ improvement has been reported by inserting an Al$_2$O$_3$ layer with open point contacts.[24-25]

Recently, we observed both $V_{oc}$ and quasi-Fermi level splitting improvement without a significant current loss in CIGSe$_2$, passivated by a continuous TiO$_2$ layer at the back contact deposited by thermal atomic layer deposition (ALD).[26] Thus, both GBG and dielectric metal oxides layers are effective to reduce back surface recombination.

Recombination not only influences the open circuit voltage or $\Delta E_F$, but also the fill factor (FF) of the solar cell via the diode factor. A small diode factor, close to 1, is desirable, because it increases FF of solar cells.[27] For a sufficiently doped semiconductor in low excitation, both radiative recombination and trap assisted Shockley-Read-Hall (SRH) recombination in the quasi-neutral region (QNR) only shift the minority carrier quasi-Fermi level, leading to a diode factor of 1.[28] A diode factor of 2 is mainly due to mid-gap deep defects and SRH recombination in the space charge region (SCR) where n ≈ p and thus both quasi-Fermi levels move.[28] Generally, the diode factor is determined from the dark current density-voltage ($J_d$-$V$) characteristic and we label this the electrical diode factor (EDF).[27, 29-30] Influenced by different recombination mechanisms, commonly the EDF is between 1 and 2. And the dominating recombination channel can be judged from the value of the EDF. In non-ideal cases, the $J_d$-$V$ characteristics can be dominated by series and shunt resistance, making it problematic to extract the diode factor.[29] To solve this problem, the illumination intensity dependent photoluminescence develops into a technology to extract reliable optical diode factor (ODF) of contactless absorbers at open circuit condition.[31-34] The relation between EDF and ODF has been discussed in earlier work[34-35]: they are equivalent in the situation when the finished device doesn't have additional recombination paths compared to the absorber alone. Because there's a very weak or even no p-n junction for contactless absorbers, the ODF is the diode factor for recombination in the QNR which implies it should have a constant value of 1 in low injection conditions. However due to metastable defects transitions, we usually observe an ODF larger than 1 in low injection condition experimentally.[7] The variable ODF in low injection condition due to metastable defects transition requires extending the theory by considering the influence of back surface recombination and of doping density. In this way, making the ODF more versatile as a contactless optical characterization method to predict the efficiency of solar cells.

We study solar cells based on CuInSe$_2$ with a bandgap around 1.0 eV, which makes it an interesting bottom cell for highly efficient tandem cells, provided the efficiency can be increased. With different back surface passivation strategies, for example by introducing Al$_2$O$_3$, TiO$_2$ or GBG, we successfully control the back surface recombination velocities and realize a $\Delta E_F$ improvement up to ~40 meV compared to the unpassivated reference sample on an Mo back contact. This allows us to study the influence of back surface recombination velocity on the ODF with an aim to eventually understand which recombination mechanism controls its value. As a result, we find back surface unpassivated samples with a lower $\Delta E_F$ exhibit a smaller ODF, indicating that back surface recombination reduces the ODF, which means that the diode factor of 1 is not always necessarily an indicator for an efficient solar cell. We study CI(G)Se$_2$ films with different doping level. These experiments show in the contrary, that CI(G)Se$_2$ with a higher net doping density has a higher $\Delta E_F$ and presents a lower ODF. Thus, a higher ODF can be the result of insufficient doping of CI(G)Se$_2$ which also contributes to loss of $\Delta E_F$ and hence $V_{oc}$. We perform SCAPS simulations based on our experiments to understand our results and to explain how back surface recombination and doping density influence the ODF. We clarify the ODF is mainly dominated by the extra change of the carrier density with change in illumination intensity, which can reveal the back surface recombination and doping level of absorbers.

## 2. Results and discussion

### 2.1 The effects of back surface recombination on $\Delta E_F$

In this work, we employ the one-dimensional modeling solar cell simulator SCAPS[36] and experiments to investigate how much $\Delta E_F$ can be improved by reducing back surface recombination of CISe$_2$ solar cell with a bandgap around 1.0 eV. All parameters used in the simulations can be found in **Table S1**. **Figure 1**a shows the quasi-Fermi level splitting as a function of back side recombination velocity $S_b$ and bulk lifetime $\tau_b$, as obtained by SCAPS simulations. The front surface recombination velocity used in these simulations is $1.4\times10^3$ cm/s, which can be realized by CdS covering.[10] For samples with a bulk lifetime ($\tau_b$) below 25 ns, a significant $\Delta E_F$ loss is observed only when $S_b > 10^4$ cm/s, because otherwise the $\Delta E_F$ is dominated by the short bulk lifetime. With a bulk lifetime more than 25 ns, a dramatic loss of $\Delta E_F$ occurs already when $S_b > 10^3$ cm/s. In a typical lifetime range of CISe$_2$ solar cells from 25 to 50 ns,[37-38] the highest $\Delta E_F$ of ~545 meV can be achieved. With more advanced deposition technologies or alkali element PDT, lifetimes of ≥200 ns have been achieved,[39-42] which leads to the highest $\Delta E_F$ of 560 meV as simulated here. For the samples presented in our experiments, the longest bulk lifetime was determined to be 37 ns (see below). With this lifetime the simulations (Figure 1b) predict a $\Delta E_F$ gain of ~ 50 meV by reducing $S_b$ from $10^6$ to $10^2$ cm/s. When the bulk lifetime is long enough to make minority carrier diffusion length comparable to absorber thickness (around 50 ns in our case with electron mobility $\mu_n = 100$ cm$^2$/Vs[43] and film thickness of 3 μm), the surface recombination becomes a more important parameter to influence $\Delta E_F$ compare to bulk life time. Compared to back surface recombination, the recombination velocity at the front surface has an even stronger influence on $\Delta E_F$ due to the most of photons are absorbed in the close front surface region. As a result, the $\Delta E_F$ stays nearly the same when the bulk lifetime is longer than 100 ns with $S_b < 10^2$ cm/s as shown in Figure 1a with a black arrow. In this case, the deficits of $\Delta E_F$ is dominated by front surface recombination. The details about influences of front surface recombination on $\Delta E_F$ were also simulated and discussed in the supporting information (**Figure S1**). It shows that another 40 meV improvement of $\Delta E_F$ can be achieved by reducing the front surface recombination velocity from $1.4\times10^3$ to $10^0$ cm/s.

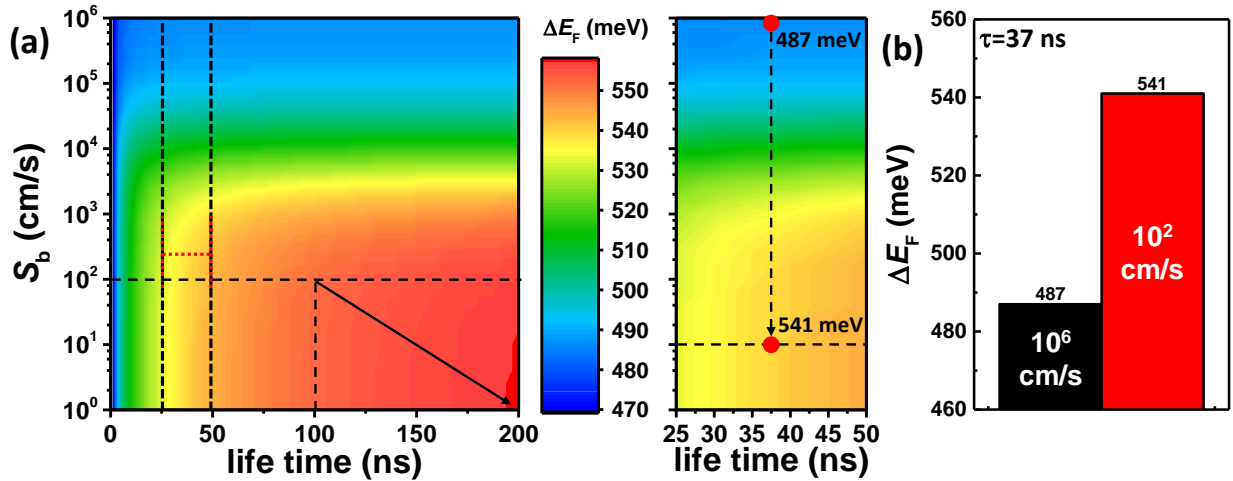

**Figure 1.** a) With doing density $N_A$ of $1\times10^{16}$ cm$^{-3}$, film thickness of 3 μm, electron mobility of 100 cm$^2$/Vs and front surface recombination velocity of $1.4\times10^3$ cm/s,[10] the SCAPS simulated $\Delta E_F$ increases with decreasing back surface recombination velocity and increasing minority carrier lifetime. Due to the limitations of front surface recombination, further decreasing $S_b$ and increasing bulk lifetime along the black arrow in the lower right corner can only slightly increase $\Delta E_F$. The second figure is an enlargement of the first figure range from 25 ns to 50 ns; b) With an approximate bulk life time of 37 ns, around 50 meV $\Delta E_F$ can be improved by reducing the back surface recombination velocity from $10^6$ cm/s to $10^2$ cm/s.

To reduce back surface recombination experimentally, both strategies of dielectric metal oxide layers and GBG are applied. The sample structures studied in experiments are shown in **Figure 2**.

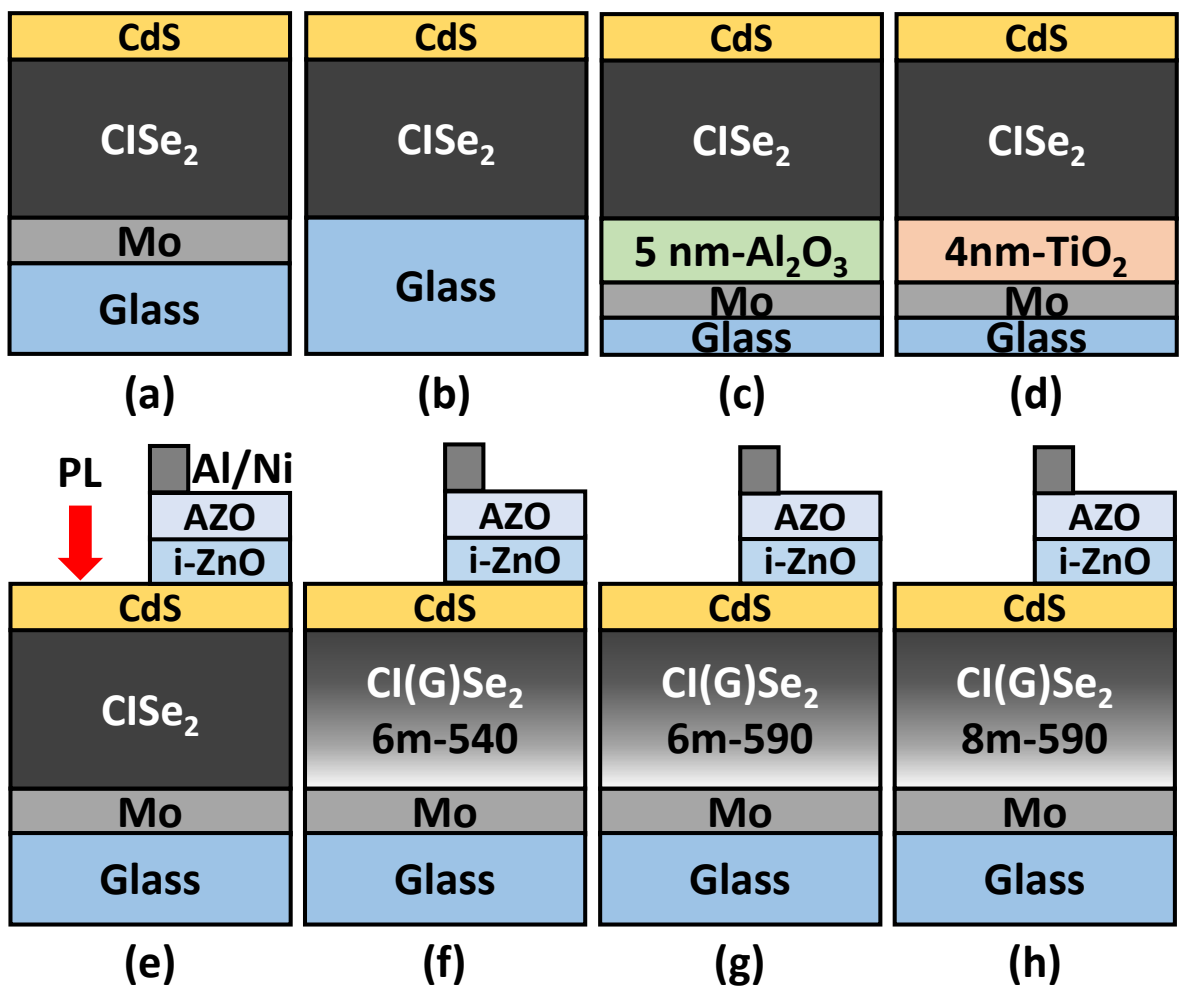

**Figure 2.** Samples to study the back surface passivation effects of dielectric layers: a) Reference CuInSe$_2$ sample grown on Mo; b) CuInSe$_2$ directly grown on soda lime glass; c) d) CuInSe$_2$ with a 5 nm Al$_2$O$_3$ or 4 nm TiO$_2$ passivation layer at back contact respectively. To study the passivation effects of gallium back gradient (GBG): e) Reference CuInSe$_2$ sample without Ga gradient; f), g), h) Sample grown with a 6/6/8 minutes CuGaSe$_2$ pre-deposition followed by a CuInSe$_2$ growth with the highest substrate temperature reaching 540/590/590 °C, respectively. Some parts of the samples are covered by sputtered i-ZnO, AZO and e-beam evaporated Al/Ni grids to finish the devices. All of the PL measurements are conducted for samples only covered with CdS.

We prepared 4 samples to study the passivation effects of dielectric layers, which all came from the same deposition process run. The first one is a reference CuInSe$_2$ grown on Mo without any back contact passivation (Figure 2a). The second one was grown directly on a clean soda-lime glass (Figure 2b). Another two dielectric layer passivated samples have a 5 nm Al$_2$O$_3$ layer (Figure 2c) and a 4 nm TiO$_2$ layer (Figure 2d) between CuInSe$_2$ and Mo, respectively. All of these samples were covered by CdS to reduce front surface recombination. Another four samples were prepared to study the passivation effects of the GBG. All of them were grown on Mo coated soda-lime glasses with passivation of the CdS layer at front surface. The reference sample (Figure 2e) is a pure CuInSe$_2$ without any Ga. The other three samples have different Ga profiles that are realized by controlling the substrate temperature and Ga supply duration. 6m-540, 6m-590 or 8m-590 means that sample experiences 6/6/8 minutes pre-deposition of CuGaSe$_2$ then followed by CuInSe$_2$ deposition with the highest substrate temperature of 540/590/590 °C as shown in Figure 2f, g, h, respectively. The GGI and corresponding $E_g$ distribution, shown in **Figure 3**, are determined by non-calibrated secondary ion mass spectrometry (SMIS) and PL. Details about how to determine GGI distribution according to these two technologies are shown in supporting information (section 3, see **Figure S2**). The left axis in Figure 3 shows the $E_g$ determined by the linear approximate dependency, $E_g^{CIGS} = E_g^{CIS} + 0.65\times(GGI)$, given in reference.[44] As shown in **Figure S3**, the parabolic function and linear approximation determined $E_g$ are very close, which makes the linear approximation is rational to be applied to plot both GGI and $E_g$ distribution along the film depth in a single figure. 8m-590 sample has the highest GGI and corresponding $E_g$ at backside because the largest amount of Ga, and that of 6m-540 sample is higher than 6m-590 sample because a lower substrate temperature reduces diffusion of Ga. The film thickness difference shown in Figure 3 is due to inhomogeneity of the film. The completed devices only made for GBG samples with sputtered i-ZnO, AZO and e-beam evaporated Al/Ni grids depositing in a row. More details about samples preparing can be found in Methods section.

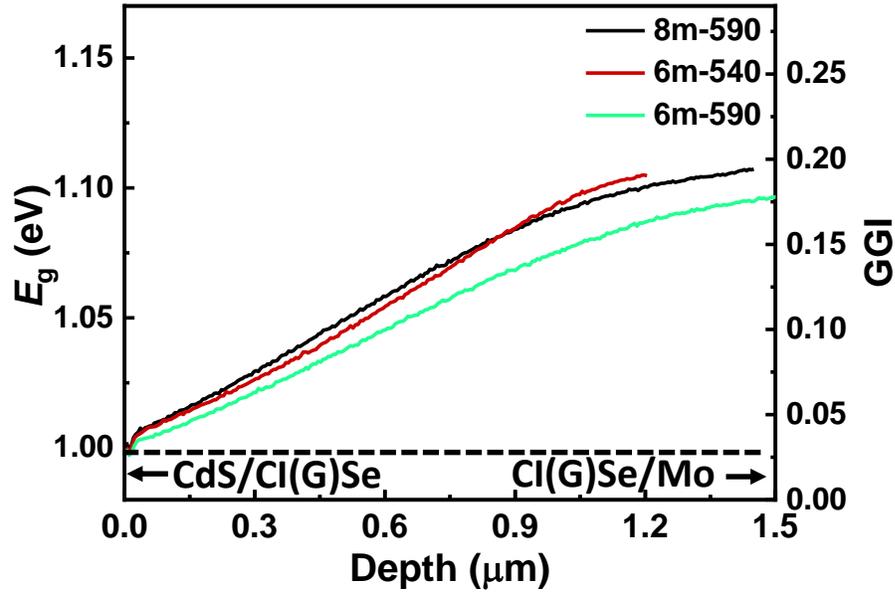

**Figure 3.** GGI and corresponding $E_g$ depth profiles for GBG samples, which are extracted from non-calibrated SIMS and PL.

**Figure 4**a shows the calibrated one-Sun PL spectra of samples without an annealing in air with dielectric metal oxide layers (top) or GBG (bottom) backside passivation. A higher radiation flux means less non-radiative recombination losses and hence a higher $\Delta E_F$. Planck's generalized law[45] determined $\Delta E_F$ are given in Figure 4b, which shows that $\Delta E_F$ of both dielectric layer and GBG passivated samples are improved compared to their reference samples. First, we discuss the results without an annealing in air. Inserting a 5-nm ALD $Al_2O_3$ layer achieves the highest $\Delta E_F$ improvement of ~40 meV that is in good agreement with our simulation prediction. This value is higher than the $\Delta E_F$ of the sample grown on glass (without Mo layer) and the sample with a 4-nm ALD $TiO_2$ layer, which are both expected to have a quite low interface recombination compared to the metal/semiconductor interface. It implies that $Al_2O_3$ is most effective in suppressing back

surface recombination among the studied passivation layers, as observed previously.[26, 46] A Similar $\Delta E_F$ gain of ~40 meV is obtained from the GBG samples, which means that the passivation

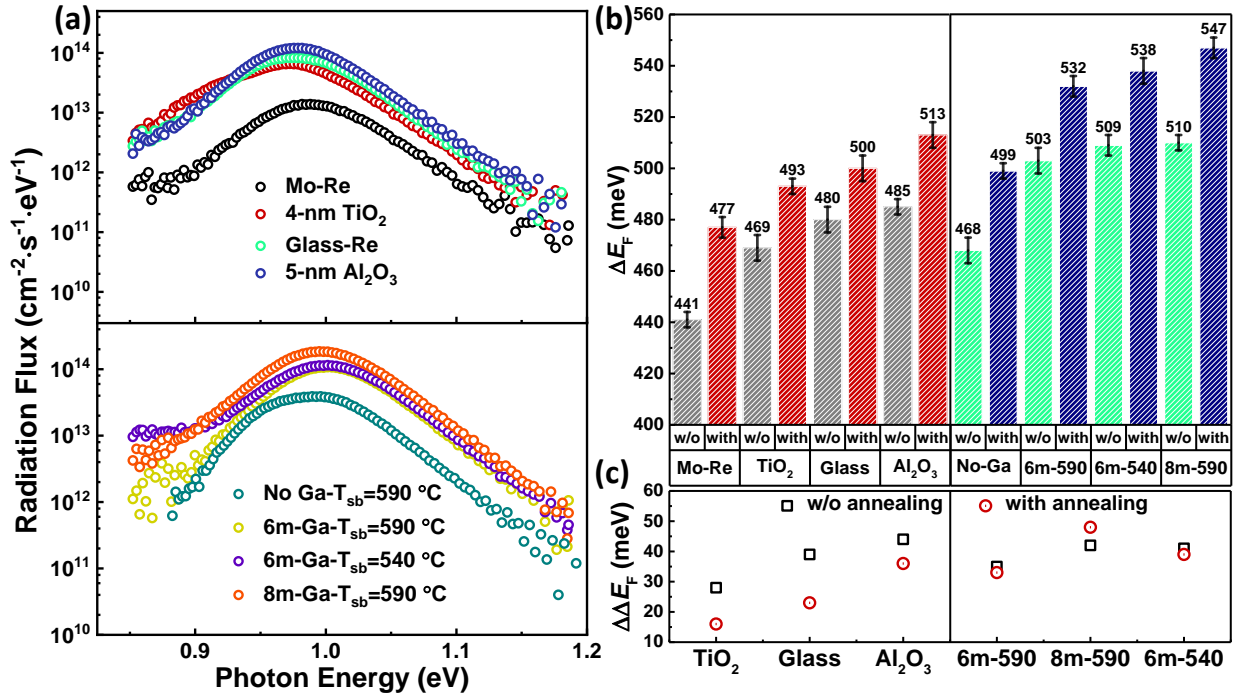

**Figure 4.** a) The calibrated PL spectra of dielectric layers and GBG passivated samples without annealing in air, measured under 1 sun equivalent illumination; b) The $\Delta E_F$ determined by Planck's generalized law are obviously improved after introducing dielectric passivation layer or Ga back gradient compared to the unpassivated reference samples, and another ~ 30 meV of $\Delta E_F$ can be gained by annealing the CdS covered samples in air at a temperature of 200 °C for 2 minutes. The "w/o" in the figure means before annealing and "with" means after annealing; c) $\Delta E_F$ improvement ($\Delta\Delta E_F$) of the passivated samples compared to the unpassivated reference sample, annealed or not annealed, respectively.

effect of GBG is comparable to $Al_2O_3$ that has been widely applied in semiconductor devices to reduce surface recombination.[21, 47-48]

To further demonstrate the passivation effects of dielectric layers and GBG, TRPL measurements were conducted to determine effective minority carrier lifetime that is described by[10]:

$$\frac{1}{\tau_\text{e}} = \frac{1}{\tau_\text{bulk}} + \frac{1}{\tau_\text{surf}} \qquad (1)$$

Where $\tau_\text{e}$ is the effective (measured) minority carrier lifetime, $\tau_\text{bulk}$ is the bulk lifetime, $\tau_\text{surf}$ is the lifetime due to front and back surface recombination. For samples with identical bulk, the effective lifetime changes with the surface recombination velocity, which means a longer $\tau_e$ indicates a lower surface recombination velocity. The TRPL decays of post-deposition annealed samples are shown in **Figure 5**a, b. Because some samples show 2-expential decay, a 2-expential fit is used to determine the weighted effective lifetimes according to Equation 8 (methods section) and results are summarized in **Table 1**. More details of 2-exponential fitting can be found in **Table S3**. The Al$_2$O$_3$ passivated sample shows the lowest decay resulting an effective minority carrier lifetime of 28.8 ns that is nearly five times longer than that of the unpassivated Mo sample. If we assume the carrier uniformly distribute, 40 meV increase in $\Delta E_F$ can be estimated according to $k_B T \cdot \ln(\frac{\tau_1}{\tau_2})$, where $k_B T$ is the thermal energy at room temperature. $\tau_1$ and $\tau_2$ are effective minority carrier lifetime with and without passivation, respectively. It is in agreement with the experimental result that the Al$_2$O$_3$ passivated sample has the highest $\Delta E_F$ of 513 meV which is ~40 meV higher than the reference sample. The longest effective minority carrier lifetime also demonstrates Al$_2$O$_3$ has better passivation effects than TiO$_2$ and glass, because they have a shorter effective minority carrier lifetime of 16.1 ns and 19.9 ns, respectively. In spite of this, the lifetimes of them are still much longer than that of the reference sample (6.0 ns), which supports again that they also have the capability to reduce interface recombination.

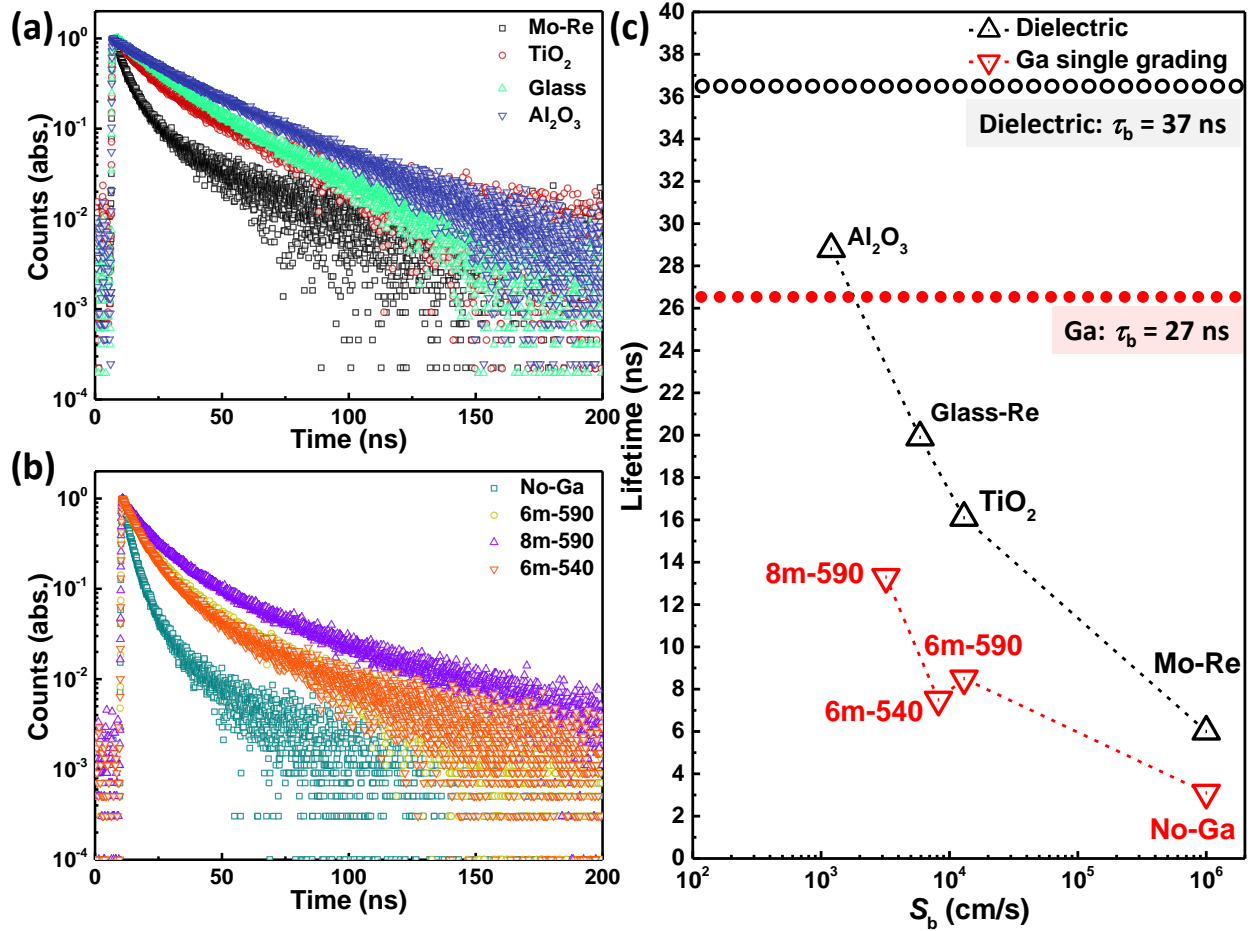

**Figure 5.** TRPL measurements of samples with post-deposition annealing in air: a) The PL counts decay of dielectric layer or glass passivated samples compared to the unpassivated sample; b) The PL counts decay of GBG samples compared to the reference sample without Ga gradient; c) The lifetime decreases with increasing back surface recombination velocity. Lifetime shown here is weighted effective lifetime determined from a fitting with a 2-exponential decay function. *Effective* back surface recombination velocities are extracted by SCAPS simulation (see Figure 6).

The lifetimes of GBG samples are also longer than their reference sample without gradient. The longest lifetime of 13.3 ns is realized by the sample with 8 minutes CuGaSe$_2$ pre-deposition and followed by CuInSe$_2$ deposition with the highest substrate temperature of 590 °C (sample No. 8m-590), which suggests it has the lowest back surface recombination. This is the sample that has the most Ga provided during growth. Further explanation comes from the GGI distribution shown in Figure 3. The 8m-590 sample shows the highest Ga distribution with the highest corresponding bandgap that is in agreement with the best passivation effects contributing to the highest $\Delta E_F$ shown in Figure 4b. In particular, a higher bandgap at the backside means a lower back surface recombination due to a lower electron density (minority carrier density). With decreasing in the Ga amount and keeping the same high substrate temperature of 590 °C, the 6m-590 sample has the lowest bandgap at the backside which results in a shorter lifetime of 8.5 ns. Reducing substrate temperature reduces the diffusion of Ga. As a result, the back side bandgap of the 6m-540 sample is very close to that of the 8m-590 sample. However, the minority carrier lifetime of 6m-540 sample is slightly shorter than 6m-590 sample. We believe that this is mainly caused by the higher substrate temperature which usually improves the quality of the absorber leading to a longer minority carrier lifetime of the bulk. [49]

Rapid low temperature post-deposition annealing of CI(G)Se devices in air was demonstrated to improve $V_{oc}$ by ~60 millivolt.[50-51] For our uncompleted devices covered only with CdS, this improvement was also confirmed after 200 °C and 2 minutes post-deposition annealing in air. Figure 4b shows that $\Delta E_F$ of the samples improved by another ~30 meV after the post-deposition annealing independent of the back surface passivation scheme. This improvement has been explained by a slight inter-diffusion of Cd or S atoms at CdS/CI(G)Se interface, which reduces the front surface recombination.[52] Front surface passivation should be detectable by an increased effective minority carrier lifetime. We studied effective minority carrier lifetime by TRPL on the samples with dielectric layers before and after annealing. Actually the only considerable lifetime improvement from 2.1 ns to 6.0 ns is observed for the reference sample after the annealing, but the lifetimes of dielectric layer passivated samples essentially stay the same (see Table S3 and **Figure S4** in the supporting information). This observation indicates that the $\Delta E_F$ improvement introduced by annealing in these samples is not due to passivation effects, but rather dominated by increase in the net doping density.[53-54] However, both lifetime and doping density improvement play a role for the reference sample, that is the reason why the reference sample achieves the highest improvement in $\Delta E_F$ after annealing (Figure 4b) and the $\Delta E_F$ improvement ($\Delta\Delta E_F$) of dielectric passivated samples ($\Delta\Delta E_F = \Delta E_F^{\text{dielectric}} - \Delta E_F^{\text{Mo}}$), shown in Figure 4c with red circles at left side, is lower after annealing.

Doping density plays a significant role on $\Delta E_F$. The GBG samples have shorter lifetimes but higher $\Delta E_F$ compare to dielectric layer passivated samples, which suggests a higher doping density of GBG samples. Doping density $N_A$ of the absorbers can be estimated from the following equation (see supporting information section 7 for the derivation):

$$N_A = p_0 = \frac{d \cdot N_c N_v}{G \cdot \tau_n^{\text{eff}}} \exp\left(\frac{\Delta E_F - E_g}{k_B T}\right) \qquad (2)$$

Where $N_c$ and $N_v$ is the effective density of states of the conduction and valence band, respectively. $\tau_n^{\text{eff}}$ is effective carrier lifetime obtained from TRPL measurements. $G$ is the generation flux, d is the thickness of the film and $k_B T$ is the thermal energy at room temperature. This is a completely contactless optical characterization method to extract doing density compare to classical capacitance-voltage (C-V) or Hall measurements. The calculated results are shown in Table 1. To calculate net doping density, we assume a homogeneous distribution of electrons, also for the GBG samples, which means we neglect the electron confinement due to a bandgap gradient. We estimate the effect of confinement in a simulation (see supporting information section 8 and **Figure S5**). The simulation results show that the confinement effect is weak: only ~10 meV of $\Delta E_F$ improvement can be achieved by introducing a conduction band gradient that is similar to the 8m-590 sample. Furthermore, the ungraded reference sample results in a very similar doping level as the GBG samples. With this, we conclude that the observed $\Delta E_F$ differences between dielectric layer and GBG passivated samples are mainly due to the different doping density. The doping density of GBG samples and their ungraded reference sample is ~ $1.8\times10^{16}$ cm$^{-3}$ that is higher than the doping level of the dielectric passivated samples of ~$8.0\times10^{15}$ cm$^{-3}$. This difference is caused by the composition difference among these absorbers, because the Cu ratio (Cu/In) of dielectric passivated sample is around 0.87 which is much lower than that of GBG samples (Cu/In+Ga) of around 0.96. It has been shown previously that a higher Cu content increases the p-type doping.[55-56] This is likely caused by a higher compensation in more Cu-poor samples due to the increased density of the shallow donor In$_{Cu}$ antisite defect which decreases the p-type doping.[57] But on the other hand, the Cu poor CI(G)Se has a longer minority carrier lifetime (the dielectric layer passivated sample series in Table 1) as reported by previous researches.[58] This partially explains why dielectric layer passivated samples have a longer minority carrier lifetime but a lower Δ$E_F$ when compared to GBG samples.

Comparing Δ$E_F$ and lifetime can only give us a general idea about passivation effects of dielectric layers and GBG. To make it more comparable, SCAPS simulation is introduced to estimate specific back surface recombination velocity for different samples. The bulk lifetime is needed to do these simulations and that can be estimated based on the reference sample. The reported back surface recombination velocity between Mo and CuInSe$_2$ ($10^6$ cm/s)[59-60] is much higher than the reported surface recombination velocity of CdS passivated front surface ($1.4\times10^3$ cm/s).[10] In this case, the following relation between surface lifetime and recombination velocity has been derived [61]:

$$\tau_s \cong \frac{d}{S_b} + \frac{4}{D}\left(\frac{d}{\pi}\right)^2 \tag{3}$$

Where $d$ is the sample thickness, $S_b$ is the back surface recombination velocity and $D$ is the diffusion constant of the minority carriers. $D$ can be determined by the Einstein relation, $D=$ $\frac{\mu_e k_B T}{q}$, where $k_B T$ is the thermal energy at room temperature, q is elementary charge and $\mu_e$ is electron mobility, $\mu_e$ = 100 cm$^2$/Vs is used for Equation 3 and SCAPS simulations.[43] Similar values of $\mu_e$ were found in Hall measurements on epitaxial n-type films.[62] The thickness of the reference samples is determined by SEM cross section shown in **Figure S6**. Then by introducing Equation 3 into Equation 1, the approximated bulk lifetimes of the reference samples for the dielectric layer and GBG passivated samples are calculated from the measured effective lifetimes $\tau_n^{eff}$. Only the bulk lifetime of the two reference samples are calculated as these samples have a Mo/CuInSe$_2$ interface and thus a $S_b$ (10$^6$ cm/s)>>$S_f$ (1.4×10$^3$) as assumed for Equation 3. The bulk lifetime of GBG and dielectric layer passivated sample is 27 ns and 37 ns, respectively. Experimentally, the bulk lifetimes are assumed to stay constant with back surface passivation approaches and Equation 3 with Equation 1 is used to determine the back surface recombination velocity.

For the SCAPS simulations, we can use the determined doping density and the bulk life time to simulate $\Delta E_F$ with respect to different back surface recombination velocities. The results are shown in **Figure 6**. Evidently, $\Delta E_F$ decreases with the increase in back surface recombination velocities. By comparing the $\Delta E_F$ acquired by experiments and SCAPS simulations, the back surface recombination velocity can be estimated and the results are summarized in Table 1. It is noted that the backside recombination velocity determined above for the GBG samples is an *effective* one: i.e. the recombination velocity that would be present at an equivalent backside with flat bands. In principle, the back surface recombination ($R_s$) is given by $R_s \approx S_b \Delta n$, $\Delta$n is the minority carrier (electron) density. The Ga gradient reduces back surface recombination by reducing $\Delta n$ rather than reducing the surface defects density or their capture cross-section that determines $S_b$. The change of $\Delta n$ in this case is determined by conduction band edge difference ($\Delta \emptyset$) at backside between no

Ga and Ga sample. Therefore, the backside recombination of grading samples ($R_s^{gra}$) can be expressed as:

$$R_s^{gra} \approx S_b \Delta n_{gra} = S_b \Delta n_{flat} \exp(\frac{-\Delta \emptyset}{k_B T}) = S_b^{eff} \Delta n_{flat} \quad (4)$$

Where $\Delta n_{gra}/\Delta n_{flat}$ is the minority carrier density of GBG/no Ga samples, $S_b^{eff}$ is an *effective* back surface recombination velocity that is equal to $S_b \exp(\frac{-\Delta \emptyset}{k_B T})$. With this transition, the passivation effects of dielectric layers and GBG can be easily compared in terms of recombination velocities.

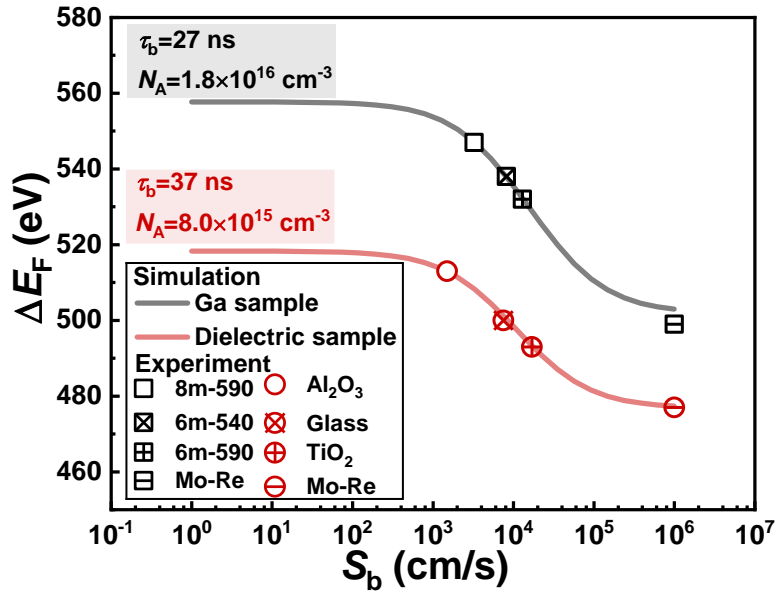

**Figure 6.** The solid lines are simulated $\Delta E_F$ changes with $S_b$. Red circles/black squares are measured $\Delta E_F$ of dielectric layers/GBG passivated samples with their unpassivated reference. The $S_b$ of the actual samples can be estimated by comparing the experimental $\Delta E_F$ to the simulated results.

For the dielectric layer passivation, Al$_2$O$_3$ gives the lowest $S_b = 1.2 \times 10^3$ cm/s due to the highest $\Delta E_F$ and longest minority carrier lifetime. Similar but slightly lower passivation effects of the glass back contact ($S_b = 5.9 \times 10^3$) is observed. The $S_b$ of the TiO$_2$ sample is higher at $1.3 \times 10^4$ cm/s due to loss of 20 meV $\Delta E_F$ compared to the Al$_2$O$_3$ sample. Better passivation effects of Al$_2$O$_3$ compared to TiO$_2$ were also reported in other works.[26, 46] All dielectric passivation layers show a considerable decrease of $S_b$ and improvement of $\Delta E_F$ compared to unpassivated Mo reference sample, demonstrating good passivation effects of dielectric layers. The Ga gradient has a similar passivation effect, where sample with the highest band gap at backside (8m-590) has the lowest *effective* $S_b$ of 3.2 ×10$^3$ cm/s that is only a bit higher than that of the Al$_2$O$_3$ sample. It demonstrates that the passivation effects of the Ga gradient is comparable to Al$_2$O$_3$. To verify the reliability of *effective* back surface recombination velocities extracted from SCAPS simulations, the *effective* back surface recombination velocities calculated from an analytical approach according to Equation S6, 7 are summarized in supporting information (section 9). The analytically solved $S_b$ values are summarized in Table S4, which shows that the $S_b$ extracted by analytical method for GBG samples are a little bit higher than SCPAS simulation results. But for dielectric layer passivated samples, these values are a little bit lower compared to SCAPS simulation results. The differences among them are small considering the approximations we make. The trends are the same as that getting from the SCAPS simulations, which confirms our conclusion on both passivation and the ODF studies, below. Additionally, the *effective* $S_b$ of GBG samples can be directly determined with $S_b^{\text{eff}} = S_b \exp(\frac{-\Delta\emptyset}{k_B T})$ by using the $E_g$ distribution shown in Figure 3. Results summarized in Table S4 confirm again the same trends. Moreover, all calculations and simulations demonstrate that the passivation effect of a GBG is very similar to dielectric layer.

In case of devices, we could not prepare working devices from dielectric layers because the current of device is blocked by these dielectric layers.[63-64] By introducing the GBG, the 6m-540 sample achieves highest efficiency of 15.7% without alkalis metal element PDT. It is clearly improved compare to the reference sample with best efficiency of 13.1% (Table 1). The lower efficiency of 8m-590 and 6m-590 sample compared to the reference sample may be caused by unoptimized deposition process of CIGSe$_2$ that results in dramatic loss of fill factor as show in **Figure S8**b, c. More results about devices can be found in supporting information section 10.

**Table 1.** Efficiency, $V_{oc}$, $\Delta E_F$, lifetime, net doping density and (*effective*) back surface recombination velocity

| Sample | Backside $E_g$ (eV) | Eff. (%) | $V_{oc}$ (mV) | $\Delta E_F$ (meV) | Lifetime (ns) | $N_A$ (cm$^{-3}$) | $S_b$ (cm/s) |
|---|---|---|---|---|---|---|---|
| 8m-590 Ga | 1.11 | 11.5 | 524 | 547 | 13.3 | 1.5×10$^{16}$ | 3.2×10$^{3*}$ |
| 6m-590 Ga | 1.08 | 10.7 | 503 | 532 | 8.5 | 1.7×10$^{16}$ | 1.3×10$^{4*}$ |
| 6m-540 Ga | 1.09 | 15.7 | 525 | 538 | 7.5 | 2.4×10$^{16}$ | 8.2×10$^{3*}$ |
| No-Ga Re | 0.98 | 13.1 | 453 | 499 | 3.4 | 1.9×10$^{16}$ | 1.0×10$^6$ |
| Al$_2$O$_3$ | | | | 513 | 28.8 | 7.8×10$^{15}$ | 1.2×10$^3$ |
| Glass | | | \ | 501 | 19.9 | 8.6×10$^{15}$ | 5.9×10$^3$ |
| TiO$_2$ | | | | 493 | 16.1 | 7.2×10$^{15}$ | 1.3×10$^4$ |
| Mo contact | | | | 477 | 6.0 | 7.6×10$^{15}$ | 1.0×10$^6$ |

*: The $S_b$ of GBG sample is an *effective* surface recombination velocity.

Back surface recombination and doping density of CI(G)Se$_2$ have strong influences on solar cells efficiency. Both dielectric layer and a GBG are effective to reduce back surface recombination thus resulting in a longer minority carrier lifetime and higher $\Delta E_F$. Minority carrier lifetime and $\Delta E_F$ are applied to estimate doping density of the absorbers. The higher doping level of GBG samples explains their higher $\Delta E_F$ compared to dielectric layer passivated samples. We further derive the specific back surface recombination velocities of these samples numerically and analytically based on aforesaid contact free optical characterization method. It allows us to study the influence of back surface recombination velocity and doping density on the optical diode factor that is a promising indicator to predict efficiency of the solar cell.

**2.2 Optical diode factor (ODF)**

*2.2.1 Background*

The diode factor is an important indicator for the performance of solar cells. Normally, a smaller diode factor (close to 1) is preferable because it results in a higher fill factor and thus improving efficiency of the solar cell. The ODF is the diode factor of the absorber alone in the absence of the pn-junction and therefore the diode factor of the quasi-natural region. It can be measured by a contactless optical method that is convenient and flexible.[35] In case of an intrinsic absorber ($n \approx p$), e.g. Perovskite[65], both quasi-Fermi levels of electrons ($E_{\text{Fn}}$) and of holes ($E_{\text{Fp}}$) are shifted with excitation, which results in the ODF of 2. For the p-type $CuInSe_2$ with a common doping density more than $10^{15}$ cm$^{-3}$ and dominated by non-radiative recombination,[66] all of $\Delta E_F$ in low injection condition comes from the shift of $E_{\text{Fn}}$, which finally results in a constant ODF of 1[33]. However in the presence of metastable defect, the ODF can be larger than 1 in low excitation condition, as recently shown by Weiss.[7] Here, we recapture the important points and critical equations that we need for a better understanding of our experiments and simulation results. The radiation flux $R_r$ from the sample is empirically found to follow a power law over many orders of generation flux (illumination intensity) $G$, $R_r \propto G^A$. Then, ODF is directly determined by the derivation of the logarithmic Radiation-Generation curve, which is described as:[7]

$$A = \frac{\text{dln}(R_r)}{\text{dln}(G)} \tag{5}$$

For doped semiconductors, following Planck's generalized law in Boltzmann approximation[45], the ODF depends on changes of the doping density $N_A$ upon illumination, which can be explained by metastable defects.[1, 7] Further, assuming that the minority carrier density $\Delta n$ increases linearly with $G$ ($\Delta n = \frac{G \cdot \tau_n^{\text{eff}}}{d}$), we get:

$$A = \frac{d \ln R_r}{d \ln G} = 1 + \frac{d \ln N_A}{d \ln G} \tag{6}$$

More details of derivation are shown in the supporting information (section 11).[7] If there are no metastable defect transitions, $\frac{d \ln N_A}{d \ln G} = 0$, which gives $A = 1$ in low injection condition. It is commonly found in CI(G)Se absorber materials, that metastable defects exist and convert from donors to acceptors upon injection of minority carriers (electrons).[67-70] The reverse process of converting metastable acceptors to donors requires thermal activation, which results in well-known persistent photoconductivity (PPC) in the CI(G)Se absorber materials.[4] One of the most notorious candidates for such a metastable defect is the $V_{se}$-$V_{Cu}$ double vacancy that can convert from shallow donor to shallow acceptor with illumination.[1] This process provides extra majority carriers under illumination, thus increasing $N_A$. This process depends on the amount of available photogenerated electrons, making $0 < \frac{d \ln N_A}{d \ln G} < 1$. Consequently, the ODF larger than 1 is obtained, even in low injection condition.[7] Obviously, Equation 6 shows that the ODF in a low injection condition is highly controlled by the change of majority carrier density with generation flux. In the flowing, we will study how the doping level influences this dependence and finally the value of the ODF. Furthermore, we will discuss the influence of back surface recombination. Previous studies show that the surface recombination velocity is injection level dependent [71-72], which means the linear assumption, $\Delta n \propto G$, is not proper for the samples with a high back surface recombination. In this case, we can not assume d ln$n$/d ln$G = 1$ as $\tau_{eff}(G)$ is a function of generation rather than a constant. The fundamental analytical equation should be applied to describe ODF of samples with high back surface recombination in low injection condition:

$$A = \frac{d \ln R_r}{d \ln G} = \frac{d \ln p}{d \ln G} + \frac{d \ln n}{d \ln G} \tag{7}$$

*2.2.2 Experiments and SCAPS simulation results*

To study the influence of back surface recombination on ODF, we measured illumination intensity dependent photoluminescence of dielectric layer passivated samples and their unpassivated reference. Results are shown in **Figure 7**a, and the ODFs are determined by Equation 5 with single power law fit that is a good approximation when the measurements are carried out only over 1-2 orders of magnitude of generation flux. This fit thus is generally applied to estimate the ODF of experiment and simulation results in one sun range. The unpassivated reference sample has the lowest ODF around 1.10. This value increases to 1.15 when introducing a TiO$_2$ passivation layer between Mo/CuInSe$_2$. And a larger value around 1.27 is obtained with Al$_2$O$_3$ or glass back contact. Therefore, it appears that back surface recombination has a tendency to reduce the ODF. To make this observation more robust, SCAPS simulations were conducted, and results are shown in Figure 7b, c. Because the doping density has an influence on the ODF which will be discussed later, we keep a value of $8.8 \times 10^{15}$ cm$^{-3}$ in these simulations that is similar to the experimentally determined value for this sample series (see Table 1). More parameters used in the SCAPS simulations can be found in Table S1 and S2. As discussed above, the ODF>1 due to metastable donors transform to acceptors in low generation flux region between ~$1 \times 10^{14}$ to ~$1 \times 10^{18}$ cm$^{-2}$s$^{-1}$ is observed and shown in Figure 7c. In the beginning, the ODF increases with illumination intensity because there are enough donors that can convert to acceptors. And after the ODF reaching the first peak at around ~$1 \times 10^{16}$ to ~$1 \times 10^{17}$ cm$^{-2}$s$^{-1}$ (depends on effective minority lifetime), it reduces with the increase in illumination intensity to ~$1 \times 10^{18}$ cm$^{-2}$s$^{-1}$ because the number of metastable defects in the donor states, available for the transformation, decrease with the increase in generation flux. When illumination intensity is beyond ~$1 \times 10^{19}$ cm$^{-2}$s$^{-1}$, an increase in the ODF towards a value of 2 indicates the transition to high excitation where the number of excited carriers is enough to shift both $E_{fn}$ and $E_{fp}$. The increase in the ODF due to metastable defects becomes flatter with an increase in back surface recombination velocity, as can be seen in Figure 7c with red and black spheres, which suggests a decrease in the ODF. Because the ODF peak introduced by metastable defect transitions in the one sun illumination range is broad enough, the simulated Radiation-Generation curve, show in Figure 7b, can be fitted by a single power law that agrees with our experimental results. The SCAPS simulation results shown in Figure 7b demonstrate that back surface recombination reduces the ODF from 1.26 to 1.12 when back surface recombination velocity increases from $10^0$ to $1\times10^5$ cm/s. The ODF simulated with a $S_b$ of $10^1$ and $10^2$ cm/s are not shown here since they are almost the same as the ODF of situation that has a $S_b$ of $10^3$ cm/s, i.e. a back surface recombination velocity of $10^3$ cm/s or lower has no influence on the ODF. The simulation gives an ODF around 1.26 with a low $S_b$ at the order of $10^3$ cm/s or below. It's nearly the same as the experimentally determined ODF of $Al_2O_3$ (1.27) and glass (1.28) sample that have a $S_b$ at the order of $10^3$ cm/s ($S_b[Al_2O_3]$= $1.2\times10^3$ cm/s, $S_b[Glass]$= $5.9\times10^3$ cm/s as summarized in Table 1). With a higher $S_b$ of $1\times10^4$ cm/s, the simulated ODF reduces to 1.16 which is similar to our experimental result of 1.15 for $TiO_2$ sample with a $S_b$ around $1.3\times10^4$ cm/s. And both experiment and simulation give us the lowest ODF around 1.11 when back surface recombination velocity reaches $10^5$ cm/s or above, which is the case of the unpassivated reference device. These results prove that back surface recombination reduces the ODF and shifts it closer to 1. Therefore, ODF close to 1 is not always a reliable indicator to show the absence of metastable defects, especially in the case of serious back surface recombination.

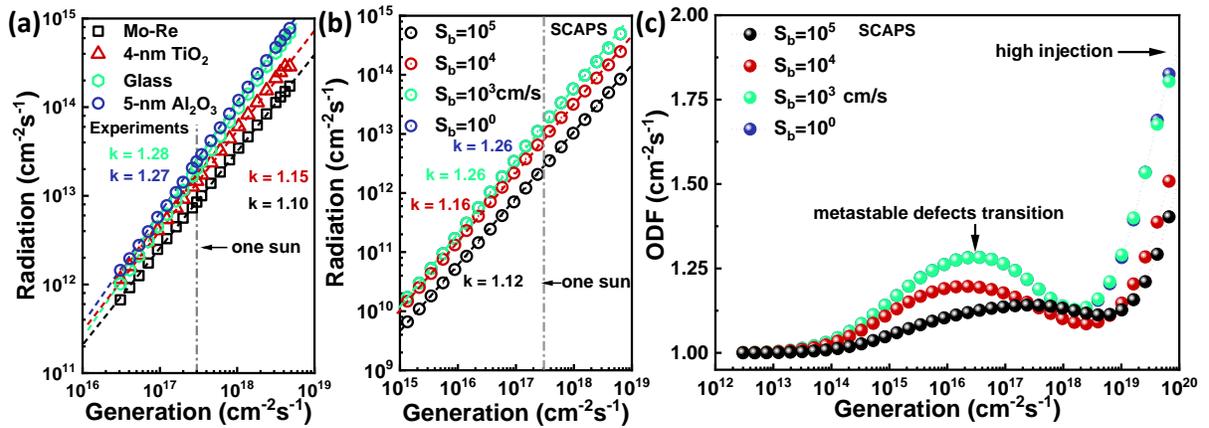

**Figure 7.** a) The ODF of unpassivated reference and dielectric layer passivated samples after annealing: the radiation flux of actual samples follows a power law with generation flux within two orders of magnitude around one sun illumination. The ODF is the exponent, as defined by Equation 5; b) The SCAPS simulated radiation-generation dependence. The curve representing $S_b = 10^0$ is not visible in (b) and (c) because it is almost the same as the curve with $S_b = 10^3$ cm/s; c) The SCAPS simulated ODF changes with the generation flux for different backside recombination velocities (net doping density $N_A = 8.8 \times 10^{15}$ cm$^{-3}$, metastable defects density $N_t$ = $8.0 \times 10^{15}$ cm$^{-3}$).

As discussed above, serious back surface recombination reduces the ODF which means samples with a low back surface recombination should have a higher ODF. However, the ODFs of all GBG samples are quite low and close to 1.1, as shown in **Figure 8**a, which is unexpected, taking their rather low $S_b$ into account. We believe this is mainly caused by the higher doping density of the GBG samples at $1.8\times10^{16}$ cm$^{-3}$ compared to dielectric layer passivated samples at $8.0\times10^{15}$ cm$^{-3}$. A higher doped sample has a higher hole density reservoir. For the same amount of metastable donors transform to acceptors, it reduces derivation of the dynamic process (d ln$p$/d ln$G$) and thus resulting in a smaller ODF. This is also clearly demonstrated by the SCAPS simulation shown in Figure 8c, where the ODF peak becomes lower and shifts to a higher injection region with increase in doping density. The simulated results of the ODFs in one sun range are directly shown in Figure 8b, it suggests the ODF decreases from 1.26 to 1.03 when the doping density increases from 8.8×10$^{15}$ cm$^{-3}$ to 3.0 ×10$^{16}$ cm$^{-3}$. In conclusion, both experiment and simulation results demonstrate that serious back surface recombination or high doping density can reduce the ODF. Therefore, it is important to be aware that a small ODF can be realized by both back surface recombination and doping density. A more serious back surface recombination reduces ODF by increasing non-radiative recombination which results in loss of $\Delta E_F$ and therefore $V_{oc}$. A higher doping density similarly also reduces the ODF but in this case with an increase in radiative recombination which causes improvement of $\Delta E_F$.

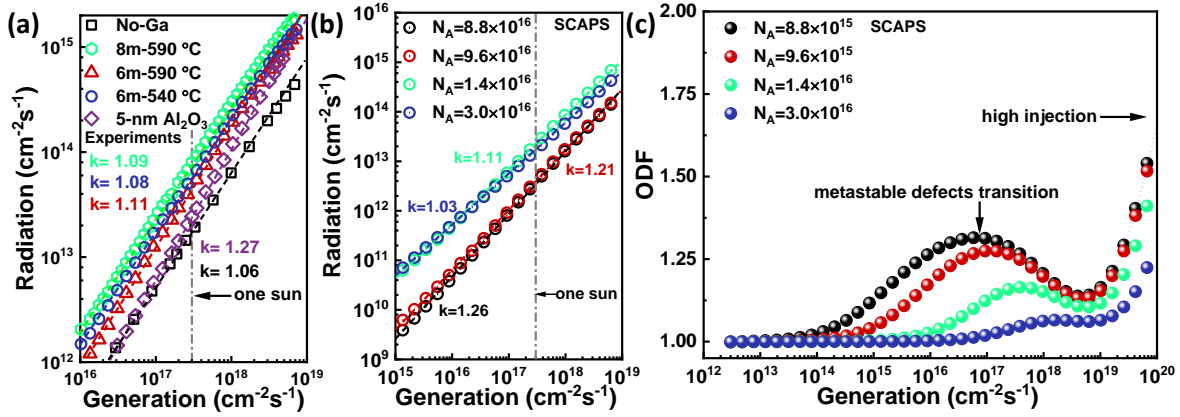

**Figure 8.** a) The ODF of GBG samples: Radiation-Generation curves of no-Ga and GBG samples with a doping density of ~1.8×10$^{16}$ cm$^{-3}$ indicate an ODF slightly larger than 1. To compare, Al$_2$O$_3$ passivated sample with a lower doping density of ~8.0×10$^{15}$ cm$^{-3}$ and a higher ODF is added; (b) The SCAPS simulated Radiation-Generation curves, which is in a good agreement to the experimental results; (c) The SCAPS simulated ODF changes with the generation flux for different net doping level, where we keep the $S_b = 1\times10^2$ cm/s.

*2.2.3 Discussion of the influence of $S_b$ and $N_A$ on ODF*

Experiment and simulation results are in good agreement and both show that a higher doping density results in a lower ODF, which is summarized in **Figure 9**a. The ODF > 1 in low injection region is due to extra hole density increased by metastable donors converting to acceptors, which consequently results in $d \ln p/d \ln G > 0$ (or $d \ln N_A/d \ln G > 0$) as shown in Figure 9b. The concentrations $n$ and $p$ considered here are from the middle of the absorber. As seen from the SCAPS simulations, shown in **Figure S10**, no matter at which position we consider the ODFs, they are essentially the same because they originate from the same mechanism that is metastable defects transition. For the same amount of metastable defects transition with different doping density, the higher doped sample has a higher majority carrier reservoir that results in a lower relative increase in the hole density compared to the lower doped sample as shown in Figure 9b with solid lines. It means $d \ln p/d \ln G$ becomes smaller in highly doped sample as shown in Figure 9b with dashed lines, which finally results in a lower ODF. The experimental hole density as a function of generation flux for two differently doped samples are also shown in Figure 9b with circles. In case of low back surface recombination, we assume that minority carrier lifetime doesn't change with different generation flux. Then we can determine the hole density with different generation fluxes from lifetime and quasi-Fermi level splitting measurements according to Equation 2. This analysis clearly shows that the hole density increases considerably by light excitation in actual samples in low injection condition due to metastable defect transitions. The GBG sample (8m-590), shown in Figure 9b by blue circles, has a higher net doping density corresponding to a smaller $d \ln p/d \ln G$, thus, a lower ODF. On the contrary, as shown in Figure 9b by black circles, the lower doped dielectric layer ($Al_2O_3$) passivated sample shows a lower hole density but a stronger increase which results in a higher ODF. In low injection condition with generation flux below $10^{17}$ cm$^{-2}$s$^{-1}$, $\Delta p$ determined by Equation S1 is more than 2 orders of magnitude lower than the $N_A$ of these samples, i.e. the sample is well inside low excitation conditions. In this case, as shown in Figure 8a, these higher doped samples have a higher radiation flux than the lower doped sample because the radiative recombination rate depends on the doping density, $R_r \approx B\Delta n N_A$. But with the increase in generation, the additional holes from metastable defect conversion reduce the difference in doping level. This effect makes the radiation fluxes from samples with different doping levels get closer with the increase in generation flux, and thus leading to a convergence behavior of the radiation-generation curves both for experimental and numerical results as shown in Figure 8a, b. The radiation fluxes getting closer with the increase in illumination intensity also indicates that the $\Delta E_F$ difference between low and high doped samples decreases (see also **Figure S9**b).

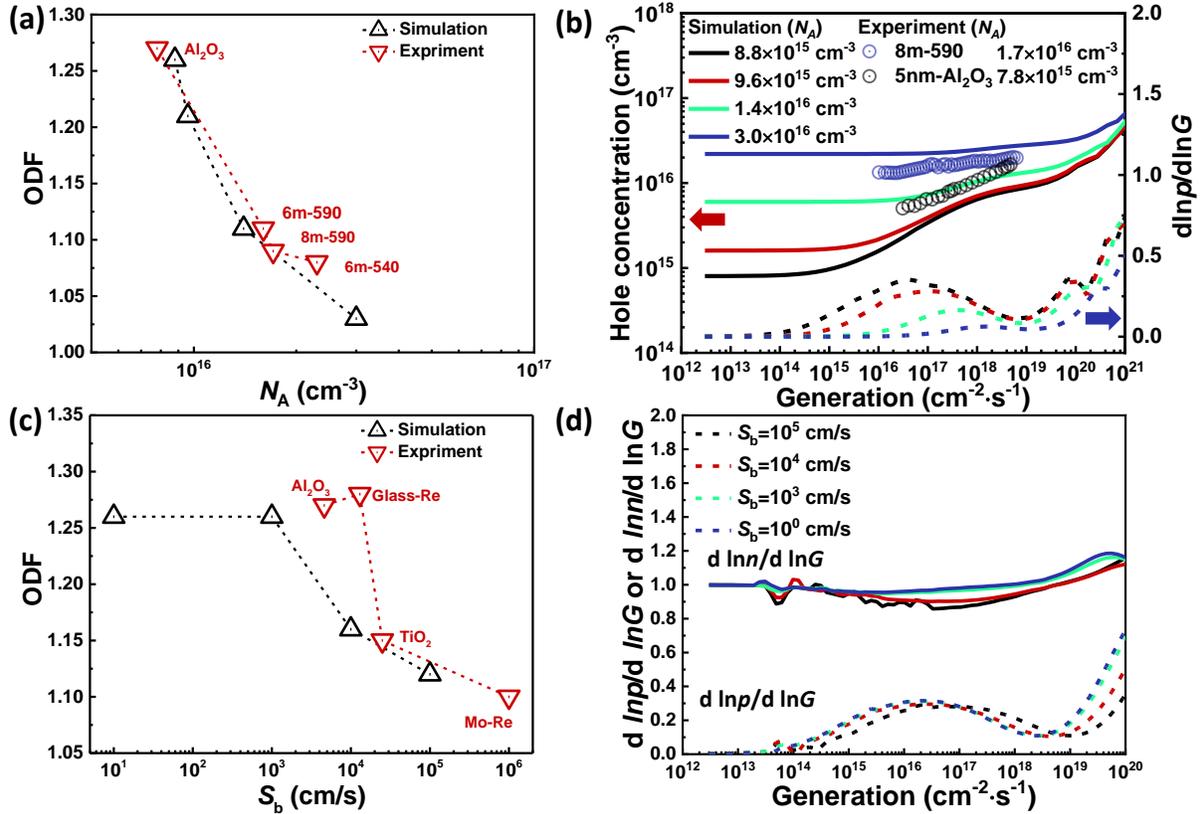

**Figure 9.** a, c) The ODF in the presence of metastable defects decreases with an increase in net doping density or in back surface recombination velocity. b) The influence of net doping density on the hole density with respect to generation in the presence of metastable defects. Solid lines are the simulation results of hole density, circles are experimentally determined values according to Equation 2. The dashed line is the simulated d $\ln p$/d $\ln G$, i.e. the ODF according to Equation 6; d) Simulation results of d $\ln n$/d $\ln G$ (solid lines) and d $\ln p$/d $\ln G$ (dashed lines) with different $S_b$. The back surface recombination increases with injection level and results in decreasing of d $\ln n$/d $\ln G$. The simulated d $\ln n$/d $\ln G$ or d $\ln p$/d $\ln G$ are taken from the middle of the absorber.

Besides the influence of the doping density, a serious back surface recombination reduces the ODF as summarized in Figure 9c. The likely explanation is given by the fact that the transformation of metastable defects from donors to acceptors depends on the availability of photo-generated electrons. If they are removed by back side recombination, the increase in the effective p-doping with the same excitation is less and thus reducing the ODF. This is confirmed by electron occupation of metastable donor states as shown in **Figure S11**. The shift of electron occupation upon generation flux explains the reason why a high $S_b$ shifts d $\ln p$/d $\ln G$ to the higher injection region but without flattening the curve as shown in Figure 9d with dished lines. Consequently, only considering the change of hole density in this case is not enough to explain why a high $S_b$ shifts and flattens the ODF at the same time as shown in Figure 7c. The ODF in this case doesn't follow the description of Equation 6 where we assume that $\Delta n$ increases linearly with generation flux ($\Delta n \propto G$), and thus d $\ln n$/d $\ln G = 1$ and consequently the ODF is only determined by the change in doping density $N_A$. The solid lines in Figure 9d show an additional loss of electron density with the increase in generation flux, which causes d $\ln n$/d $\ln G < 1$ and thus the smaller and flatter ODF according to Equation 7 (see also Figure S9a, where we show the ODF as the sum of the generation dependence of electron and hole concentration). This observation indicates that serious back surface recombination causes additional electron losses with the increase in generation flux. From SCAPS simulations, we further confirm this additional electron loss in the bulk is due to illumination intensity dependent back surface recombination that is controlled by illumination intensity determined metastable defects transition. Due to increase in the back surface recombination with increasing illumination intensity, additional electrons diffuse to the backside to compensate this additional back surface recombination upon illumination intensity, which leads to a decrease in electron density as shown in Figure 9d. This illumination intensity dependent back surface recombination indicated by SCAPS simulations originates from metastable defects transition rather than band bending at the backside in our SCAPS set-up, which is caused by the flat-band definition of the back contact in the simulation. In a simulation without metastable defects, no decrease of the electron concentration with increasing illumination is observed. All simulations and detailed discussions are shown in supporting information section 14.

In contrast to the case of samples with different doping levels, the samples with different back surface recombination velocities show a divergence behavior of radiation-generation curves upon generation flux, as shown in Figure 7a, c. In low injection, the radiation flux difference among samples are small due to rather low free electron density. This difference is getting larger and larger with the increase in illumination intensity, because the sample with a higher $S_b$ is involved in a higher extra non-radiative recombination loss that results in a lower ODF. This is also reflected by their $\Delta E_F$ difference gradually getting larger with the increase in generation flux (see Figure S9c). Therefore, the divergence behavior of radiation-generation curves may imply that the ODF differences among the samples are due to differences in back surface recombination. And the lower ODF is, the more serious back surface recombination is. This is a significant point to reveal the dominating recombination channel in solar cells, because these observations suggest that not only the value of ODF, but the behavior of radiation-generation curves can help us distinguish whether a smaller ODF originates from a higher doping level that is beneficial to efficiency or from increased (back) surface recombination that is detrimental.

## 3. Conclusion

In this work, we use photoluminescence and its generation dependence to study the exact recombination mechanisms in absorbers of solar cells, using CI(G)Se$_2$ solar cells as an example. These optical methods allow us to explore the performance of solar cells without finishing the whole devices, making it sufficiently flexible and responsive. The calibrated photoluminescence gives us an idea about $\Delta E_F$ under one sun that is the upper limit of $V_{oc}$. First, we investigate the effect of back side passivation on $\Delta E_F$. Both Ga back grading and dielectric metal oxides (Al$_2$O$_3$ as the best) passivation layer increase $\Delta E_F$ by around 40 meV compared to their respective reference samples without back surface passivation, which suggests the passivation effects of Ga gradient is as efficient as dielectric metal oxides layer. It is important to note that even for thick samples with thickness of 2 μm, back surface recombination has a significant influence on $\Delta E_F$. We can also determine the lifetime $\tau_e$ of contactless absorbers by TRPL. The longer $\tau_e$ of back surface passivated samples further demonstrates the decent passivation effects of Ga gradient and dielectric metal oxides. Because of a lower net doping density of dielectric layer passivated samples, they have a longer $\tau_e$ but a lower $\Delta E_F$ compared to Ga gradient samples. Assuming a uniform carrier distribution along the depth of the absorber, the effective net doping level of semiconductors can be calculated with $\tau_e$ and $\Delta E_F$, offering an optical method without requiring any contacts compared to traditional electrical measurements to determine doping level. Both simulation and analytical analysis can be utilized to determine surface recombination velocities, which makes it possible to directly compare passivation effects of different materials through recombination velocities. This optical method developed to determine surface recombination velocity of CI(G)Se can be extended to any solar cell materials, it gives a deeper understanding of influence of surface recombination on devices' performance.

Besides determining $\Delta E_F$, $\tau_e$ and surface recombination velocity via optical methods, the illumination intensity dependent PL gives us the ODF that is the diode factor of a contactless absorber. Normally, a smaller diode factor is preferable due to a higher fill factor. But both experiment and simulation results show that surface recombination reduces the ODF, in this case, a smaller ODF is undesirable and detrimental to the performance of solar cells. On the other hand, a higher doping level of an absorber can also reduce the ODF, which is additionally beneficial to improve $\Delta E_F$ of solar cells and hence performance. Therefore both positive and negative influences on devices' performance are able to reduce the ODF. Combining $\Delta E_F$ and the behavior of radiation-generation curve can help us distinguish recombination channels underlying the ODF. For a sample with a lower $\Delta E_F$ under one sun, compared to a reference sample, and a divergence behavior of the radiation-generation curve, i.e. increasing $\Delta\Delta E_F$, back surface recombination likely plays a critical role to limit $\Delta E_F$. A convergence behavior of the radiation-generation curves suggests that net doping density differences are more important for differences in $\Delta E_F$. Overall, with both experiments and simulations, we acquire a better comprehension of ODFs and provide particular insights into the underlying recombination processes in doped semiconductors.

## 4. Methods

### 4.1 Sample preparing

Except the sample directly grown on clean soda lime glass, preparation of all other samples begins with sputtered Molybdenum (Mo) coated soda lime glass, the thickness of Mo layer is around 500 nm. For dielectric metal oxides layer passivated sample, a TiO$_2$ or Al$_2$O$_3$ was covered on Mo layer by means of atomic layer deposition (ALD) processes. The Al$_2$O$_3$ layers were deposited by plasma assisted ALD at 200°C substrate temperature using trimethyl aluminum (TMA) as aluminum precursor and an inductively coupled remote oxygen plasma for oxidation. The growth rate of the applied plasma-assisted ALD process is 1.2 Å/cycle. With 42 cycles, it corresponds to Al$_2$O$_3$ layer thicknesses of 5 nm. The TiO$_x$ layers were deposited using thermal ALD (FlexAl reactor, Oxford Instruments), again at substrate temperature of 200°C. Tetrakis (dimethylamino) titanium (TDMAT), H$_2$O and N$_2$ are used as titanium precursor, oxidant and purge gases, respectively. The layer thicknesses as determined from the growth rate of 0.43 Å/cycle and 100 cycles amounts to 4 nm.

Dielectric metal oxide layer sample series: Polycrystalline CuInSe$_2$ thin films with thickness of ~2.0 μm were grown in a molecular beam epitaxy system by a one-stage process where the Cu, In and Se molecular flux was supplied at the same time. The set substrate temperature during the absorber growth is 550 °C for the first 15 minutes, and increases to 590 °C for the rest of the process (see also **Figure S15**a). The sample directly grown on glass or on Mo experienced the same deposition process as the dielectric layer passivated samples. Cu/In ratio of these samples, determined from energy dispersive X-ray spectroscopy (EDX), is around 0.87.

GBG sample series: Polycrystalline CuIn(Ga)Se$_2$ thin films were prepared with one-stage process, 6 or 8 minutes CuGaSe$_2$ was deposited before CuInSe$_2$ deposition (see also Figure S15b). Three different samples were prepared: 6m-590/8m-590/6m-540 sample means 6/8/6 minutes pre-deposition of CuGaSe$_2$ with substrate temperatures of 550/550/500 °C. Then followed by 15 minutes CuInSe$_2$ deposition at the same temperature and 55 minutes CuInSe$_2$ deposition at a higher substrate temperature of 590/590/540 °C. Regarding of the reference sample, it has the same deposition process to high substrate temperature prepared samples (550/590°C) but without pre-deposition of CuGaSe$_2$. The thickness of these samples is between 1.2 μm to 1.5 μm and EDX determined Cu/(In+Ga) ratio of 0.96. Changing Ga-step duration and substrate temperature leads to different Ga distribution.

After the absorber deposition, all the samples were chemical etched with 5% aqueous KCN solution for 30 s to remove potential residual oxides.[73] After etching, a chemical bath deposited (CBD) CdS buffer was applied to cover the absorber surface. The standard CBD recipe is 6-7 min deposition at 67 °C with 2 mM CdSO$_4$, 50 mM thiourea, 1.5 M NH$_4$OH. The estimated thickness is 40-50 nm according to typical growth rates. The CdS is necessary to passivate the front surface and prevent the surface degradation during the PL characterization.[40, 74-75] Additionally, a thermal PDT was performed to further improve the $\Delta E_F$, by placing the CdS covered samples on the 200 °C preheated hot plate for 2 minutes in air, followed by a natural cool down on a room temperature plate.

Devices: Because the dielectric layers block the transport of carriers which results in a rather low short-circuit current density ($J_{sc}$),[63-64] devices were only made for GBG samples. To complete the devices, a sputtered double layer of i-ZnO/AZO and e-beam evaporated Ni/Al electrodes were deposited in sequence.

**4.2 Characterization**

Absolute calibrated PL: The PL system is a home built set-up. All samples were excited with 660 nm wavelength light of diode laser with a spot diameter of ~2.6 mm in air at room temperature. The emitted photoluminescence was collected by two parabolic mirrors and then redirected to a monochromator before transmitting to an InGaAs detector with a 550 μm optical fiber. The background of PL spectra used for the $\Delta E_F$ and ODF determination were removed by a known spectrum of a calibrated halogen lamp. Quantification of both excitation and radiation flux was done by a power meter, by which we can calculate $\Delta E_F$ with specific illumination intensities from 0.01 sun, even lower, to dozens of sun equivalents, depending on the quality and $E_g$ of absorbers. The one sun means the photon flux equals to that given by AM1.5 spectrum based on $E_g$ of the absorber. All $\Delta E_F$ were measured and calculated at room temperature. According to the Planck's generalized law with the Boltzmann approximation,[45] the $\Delta E_F$ can be calculated by fitting high energy wing of the PL spectrum where we assume the absorptivity equal to 1 (a (E) = 1) with a fixed temperature of 296 K. More details about calculation can be found in other works.[75-78]

Time resolved PL (TRPL): This technology is based on time-correlated single photon counting (TCSPC) which is used to measure luminescence decays in the time domain. Measurements were taken with a 640 nm pulsed diode laser. Because some of our samples do not fully follow the 1-exponential decay, the weighted effective life time is adopted, which is calculated via:

$$\tau_e = \frac{A_1 \tau_1 + A_2 \tau_2}{A_1 + A_2} \tag{8}$$

Where $A_1$ and $A_2$ is the prefactor for the $\tau_1$ and the $\tau_2$ decay, respectively. Where possible, we also fitted lifetimes with a 1-exponential decay function and results are also summarized in Table S3. Because the difference of the single exponential estimated lifetime compared to the weighted effective lifetime is negligible and the latter can also be determined from the samples that shows 2-exponential exponential decay, the weighted effective lifetime is used here for the discussions.

Illuminated and dark current-voltage: Measurements of complete solar cells were carried out at 25 °C in a 4-probe configuration under a class AAA solar simulator that supplies a simulated

AM1.5G spectrum calibrated by a Si reference cell. The forward scanning voltage is applied from -0.3 V to 0.6 V with a step of 0.01 V.

Secondary Ion Mass Spectrometry (SIMS) depth profiles: Measurements were performed with a CAMECA SC-ultra instrument (Ametek). 1 keV focused $Cs^+$ ion beam (5 nA) was applied to sputter over a surface of the sample with an area of 250 nm × 250 nm. Only ions from the center with an area of 60 $nm^2$ were detected as $MCs^+$ or $MCs^{2+}$ where M stands for interested ions such as Cu, In, Ga, Se and Mo.

Energy dispersive X-ray spectroscopy (EDX): EDX was introduced to determine the over-all composition of the CI(G)Se$_2$ sample with an electron acceleration voltage of 20 kV. To obtain quantified results, the spectrum of each element was calibrated by their standard spectrum measured with the same electron acceleration voltage.

**4.3 SCAPS simulation setups**

SCAPS[36] is designed for the device simulation, but it is possible to use SCAPS to study $\Delta E_F$ and ODF of a single semiconductor layer. The $\Delta E_F$ can be extracted by considering the difference between electron and hole quasi-Fermi level ($\Delta E_F = E_F^n - E_F^p$). The ODF simulation is conducted by changing the neutral density (ND) setting which can give illumination or generation flux from $1 \times 10^{12}$ $cm^{-2}s^{-1}$ to $1 \times 10^{21}$ $cm^{-2}s^{-1}$.

SCAPS does not directly give us the radiation flux. But if we equal the number of photons emitted via radiation to the amount of electrons that recombine radiatively, the radiation flux can be calculated by the radiative recombination current density, $R_r = \frac{J_r}{q}$, where $R_r$ is the radiation flux, $J_r$ is the radiation recombination current density and q is the elementary charge. All parameters used in these simulations are state-of-art values that can be found in other works[7] and are summarized in the supplementary (Table S1 and S2). And it's worth to mention several specific settings:

1. The sample structure is: front contact/1 nm defect-free $CuInSe_2$/1 or 3 μm $CuInSe_2$/1 nm defect-free $CuInSe_2$/back contact. The thin defects-free layers at both sides are needed, because any contact definition generates a band bending when there are charged defects. This band bending is an artefact of our use of a device simulator for PL simulations and can be removed by these thin neutral layers. Standard is 3 μm thick $CuInSe_2$. The 1 μm $CuInSe_2$ is only used to simulate the influence of back surface recombination on the ODF, because the thinner $CuInSe_2$ shows a more obvious effect.

2. The *effective* radiative recombination coefficient in this work is found by adjusting the radiative recombination flux equals to generation flux under the one sun in the case where we keep the radiative recombination the only recombination channel (i.e. all surface recombination velocities = 0 and no defects) and make $\Delta E_F$ equals to $\Delta E_F^{SQ}$[79]:

$$R_r = \frac{J_r}{q} = G = d \cdot B n_i^2 \exp\left(\frac{\Delta E_F^{SQ}}{k_B T}\right) \tag{9}$$

Where $G$ is the generation flux, $d$ is the thickness of the absorber, $B$ is the *effective* radiative recombination coefficient, $k_B T$ is the thermal energy and $n_i$ is the intrinsic charge carrier density. As a result, the SCAPS simulator gives us a thickness dependent *effective* radiative recombination coefficient, as shown in Table S1, that is nearly 3 orders of magnitude smaller than the actual radiative recombination coefficient.[10, 80] The reason behind this is that the model used here does not satisfy the SQ model assumption with a step function of absorptivity, infinite carrier mobility or zero absorber thickness. Another important reason for the smaller *effective* B is that the radiation flux given by SCAPS is an internal radiation flux without considering the effects of light outcoupling and photon recycling. The photon recycling occurs because the interface only allows photons within the escape cone to be emitted, the rest of them are reflected back into the absorber and absorbed again. With this effect, the internal radiation flux we gain from SCAPS simulator is a factor of $4n^2$ lager than external radiation flux, that we should consider, where n is the refractive index.[81] As a result, the $\Delta E_F^{SQ}$ would be achieved with a higher radiative recombination current. To take both effects into account, the deviation from the SQ model and the difference between internal and external photon flux, we use this smaller *effective B*.

3. We include metastable defects with a density of $8\times10^{15}$ cm$^{-3}$ that is comparable to the net doping density which changes from $8.8\times10^{15}$ cm$^{-3}$ to $3\times10^{16}$ cm$^{-3}$. This is especially important for the metastable transition theory discussed in section 2.2, because only when the amount of holes gained from metastable defects converting is comparable to the net doping density, it is possible to observe the ODF larger than 1 in low injection region. For metastable defects setting, the energetic position of the donor state is at the middle of the band gap, and the acceptor state is located at 0.2 eV above the valence band edge. Setting both of them as shallow defects always results in convergence problems in our simulations. It also very important to know that these metastable configurations only work when considering equilibrium of the absorber for each illumination intensity. It means in "batch" set-up of SCAPS, the measurement working point and initial working point of different illumination intensities should be selected and simulated at the same time.

4. There is no interface defects set between defect-free CuInSe$_2$ and CuInSe$_2$, the surface recombination velocity is modified by directly changing the surface recombination velocity setting in contacts.

**Supporting Information**

Supporting Information is available from the Wiley Online Library or from the author.

**Acknowledgements**

This work was supported by the Luxembourg National Research Fund (FNR) through the PACE project under the grant number PRIDE17/12246511/PACE and through the SeVac project (C17/MS/11655733/SeVac). For the purpose of open access, the author has applied a Creative Commons Attribution 4.0 International (CC BY 4.0) license to any Author Accepted Manuscript version arising from this submission.

**Conflict of Interest**

The authors declare no conflict of interest

**Data Availability Statement**

The data used to support this study are fully open and uploaded on Zenodo. The DOI for the open data is https://doi.org/10.5281/zenodo.5813052. The data can also be obtained directly from the author.

**Keywords**

CI(G)Se$_2$ solar cell, surface recombination, quasi-Fermi Level splitting, diode factor

Received: ((will be filled in by the editorial staff))
Revised: ((will be filled in by the editorial staff))
Published online: ((will be filled in by the editorial staff))

**Supporting Information**

# Solar cell efficiency, diode factor and interface recombination: insights from photoluminescence

*Taowen Wang[*], Florian Ehre, Thomas Paul Weiss, Boris Veith-Wolf, Valeriya Titova, Nathalie Valle, Michele Melchiorr, Jan Schmidt, and Susanne Siebentritt*

1. Parameters used for simulation and calculation.

**Table S1.** Critical parameters of CuInSe$_2$ used in SCAPS simulation and calculation. The 1 μm CuInSe$_2$ is only applied to simulate the influence of back surface recombination on the ODF.

| Parameters | | Defects-free CuInSe$_2$ | CuInSe$_2$ | | |
|---|---|---|---|---|---|
| Thickness | $d$ (μm) | 0.001 | 1 or 3 | | |
| Band gap | $E_g$ (eV) | 1.02 | | | |
| Electron affinity | $\chi$ (eV) | 4.5 | | | |
| Dielectric permittivity | $\varepsilon$ (relative) | 13.6 | | | |
| Effective density of states in the conduction band | $N_c$ (cm$^{-3}$) | 7.78×10$^{17}$ | | | |
| Effective density of states in the valence band | $N_v$ (cm$^{-3}$) | 2.1×10$^{19}$ | | | |
| Electron thermal velocity | e-$V_{th}$ (cm/s) | 1×10$^7$ | | | |
| Hole thermal velocity | h-$V_{th}$ (cm/s) | 1×10$^7$ | | | |
| Electron mobility | e-mobility (cm$^2$/Vs) | 100 | | | |
| Hole mobility | h-mobility (cm$^2$/Vs) | 25 | | | |
| Shallow acceptor density | $N_A$ (cm$^{-3}$) | 8.8×10$^{15}$ - 3×10$^{16}$ | | | |
| Shallow donor density | $N_D$ (cm$^{-3}$) | / | | | |
| Effective recombination coefficient | $B$ (cm$^3$/s) | / | 1.4×10$^{-13}$ (1 um) | 4.3×10$^{-13}$ (3 um) | |
| Defect type | | / | Single donor(0/+) | Single acceptor(0/-) | neutral |
| Eletron and hole capture cross section | $\sigma$ (cm$^2$) | / | 1×10$^{-15}$ | | |

| Defects density | $N_t$ (cm$^{-3}$) | / | 8×10$^{15}$ (Metastable defects) | | 1×10$^{15}$ |
|---|---|---|---|---|---|
| Defects position | | / | Middle gap | Above $E_v$ 0.2 eV | Middle gap |
| Defects distribution | $E_t$ distribution | / | Uniform | | |
| Surface condition | | | | | |
| Parameters | | Back surface | | Front surface | |
| Surface recombination velocity of electron | $S_n$ (cm/s) | 1×10$^0$ -1×10$^6$ | | 1.4×10$^3$ | |
| Surface recombination velocity of hole | $S_p$ (cm/s) | keep the same with $S_n$ | | 1.4×10$^3$ | |
| Work function | $W$ (eV) | Auto-flat band | | | |

Table S2. Metastable defects information

| Name | Individual capture/emission processes | Expression | Energy barrier | Type |
|---|---|---|---|---|
| 'Electron capture' | Electron capture hole emission | $\tau_{EC}^{-1} = \frac{1}{\nu_{ph}} c_n c_p n N_v \exp\left(-\frac{\Delta E_{EC}}{k_b T}\right)$ | $\Delta E_{EC}$ = 0.40 $eV$ | donor-to-acceptor |
| 'Hole emission' | Double hole emission | $\tau_{HE}^{-1} = \frac{1}{\nu_{ph}} c_p^2 N_v^2 \exp\left(-\frac{\Delta E_{HE}}{k_b T}\right)$ | $\Delta E_{HE}$ = 0.81 $eV$ | donor-to-acceptor |
| 'Hole capture' | Double hole capture | $\tau_{HC}^{-1} = \frac{1}{\nu_{ph}} c_p^2 p^2 \exp\left(-\frac{\Delta E_{HC}}{k_B T}\right)$ | $\Delta E_{HC}$ = 0.35 $eV$ | acceptor-to-dononr |
| 'Electron emission' | Electron emission hole capture | $\tau_{EE}^{-1} = \frac{1}{\nu_{ph}} c_n c_p p N_C \exp\left(-\frac{\Delta E_{EE}}{k_B T}\right)$ | $\Delta E_{EE}$ = 0.94 $eV$ | acceptor-to-donor |

2. The influence of front surface recombination velocity on $\Delta E_F$.

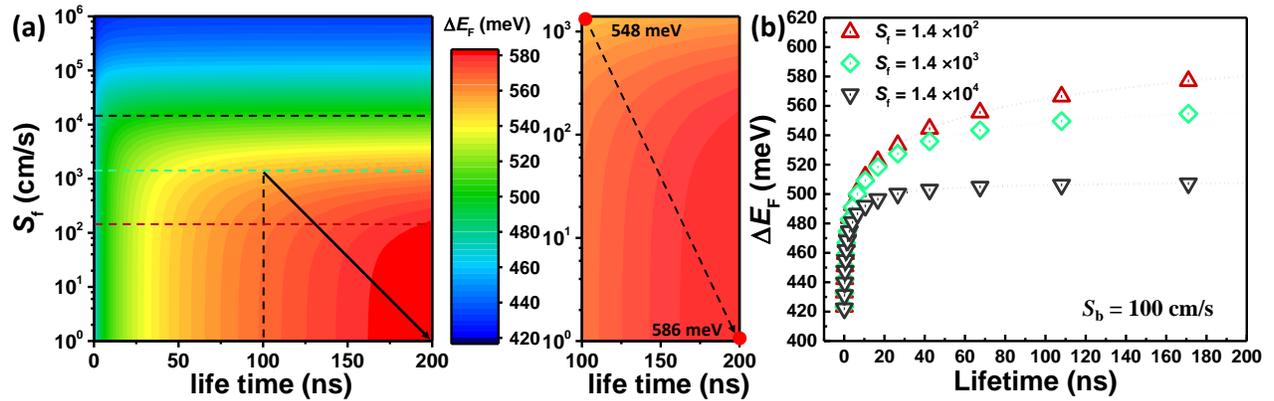

**Figure S1.** a) With a doing density $N_A$ of $1\times10^{16}$ cm$^{-3}$, film thickness of 3 μm, electron mobility of 100 cm$^2$V$^{-1}$s$^{-1}$ and back surface recombination velocity ($S_b$) of 100 cm/s, the SCAPS simulation shows another increase of around 40 meV in $\Delta E_F$ can be achieved by reducing front surface recombination velocity ($S_f$) and increasing the lifetime. The second figure is an enlargement of the first figure with lifetime range from 100 ns to 200 ns and $S_f$ range from $1.4\times10^3$ to $10^0$ cm/s, which more clearly shows this improvement is from 548 meV to 586 meV; b) Corresponding to dash lines with same the color in Figure S1a, it shows that with a low $S_b$ = 100 cm/s and long lifetime more than 100 ns, the $\Delta E_F$ is dominated by $S_f$. And with our experimental lifetime of 37 ns, further decrease the assumed $S_f$ to 1 order of magnitude doesn't make much difference as shown between green and Red curves. But black curve suggests there's a non-negligible decrease of $\Delta E_F$ around 25 meV when increasing $S_f$ to 1 order of magnitude higher.

3. Quantification of non-calibrated SIMS

The energy corresponds to the maximum of the PL spectra (Figure S2a) can be interpreted as the band gap. This PL determined band gap ($E_g^{PL}$) is the minimum band gap at front side of the absorber where the lowest Ga concentration is measured by non-calibrated SIMS (see Figure S2b, c, d). According to the dependency, $E_g^{CIGS} = E_g^{CIS}+0.65(GGI)$, given in Ref,[44] the $GGI_{PL}^{Front}$ at front side can be calculated based on $E_g^{PL}$. Comparing $GGI_p^{Front}$ to that measured by non-calibrated SIMS ($GGI_{SIMS}^{Front}$), a quantification factor $\alpha = GGI_{PL}^{Front}/GGI_{SIMS}^{Front}$ can be applied to quantify the non-calibrated SIMS GGI profile to determine the calibrated profiles of Figure 3.

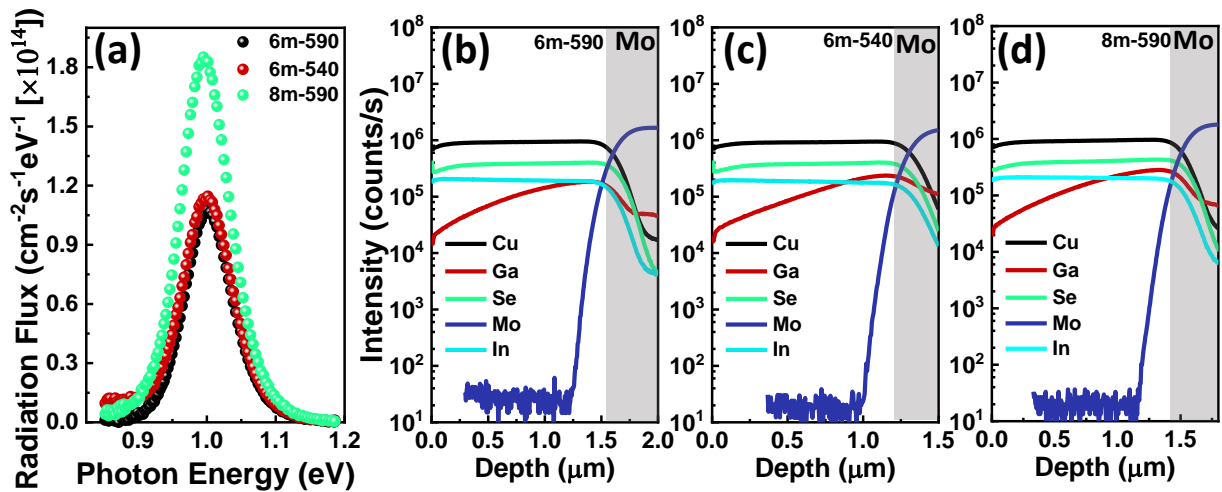

**Figure S2.** a) The PL emission of GBG samples. The PL determined band gap comes from the band gap minimum of the film that corresponding to the lowest Ga component at the front side; b) - d) The non-calibrated SIMS results of 6m-590, 6m-540 and 8m-590 sample respectively

4. The difference between linear and parabolic function determined $E_g$ distribution.

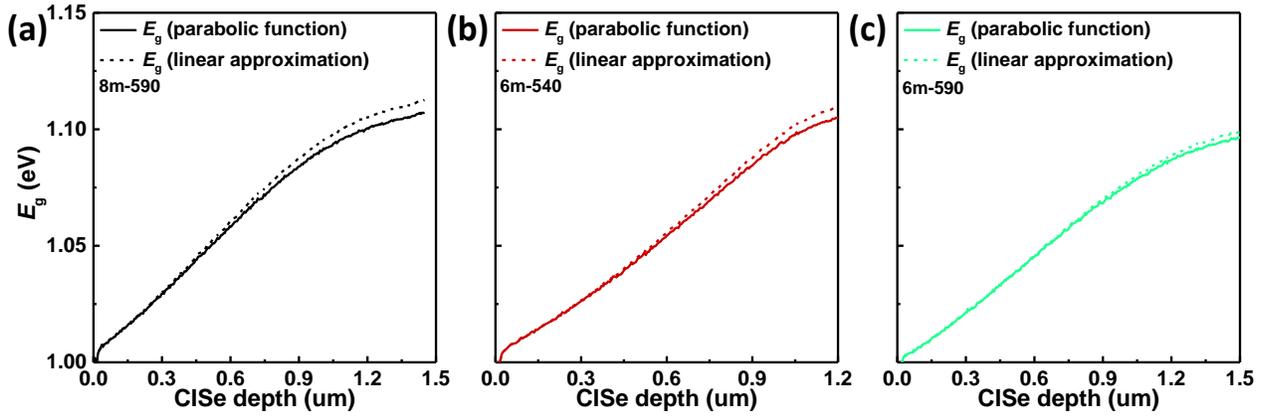

**Figure S3.** The $E_g$ distribution determined by parabolic function ($E_g^{CIGS} = E_g^{CGS}x + E_g^{CIS}(1-x) - 0.12x(1-x)$,[44] x is the Ga/(Ga+In)) is very similar to linear approximation That is the reason why we can use linear approximation to substitute parabolic function and plot both GGI and $E_g$ distribution along the film thickness in a single figure.

5. The minority carrier lifetime of samples determined by one or two exponential decay function.

Table S3. The details for 2 exponential decay fitting

| Sample | $A_1$ | $A_2$ | $\tau_1$ (ns) 2-ExpDec | $\tau_2$ (ns) 2-ExpDec | $\tau_e^*$ (ns) 2-ExpDec | $\tau_e$ (ns) 1-ExpDec |
|---|---|---|---|---|---|---|
| 6m-540 Ga (with) | 2.214 | 0.124 | 6.3 | 29.5 | 7.5 | / |
| 6m-590 Ga (with) | 1.836 | 0.135 | 6.9 | 30.6 | 8.5 | / |
| 8m-590 Ga (with) | 1.136 | 0.140 | 9.7 | 42.7 | 13.3 | / |
| Mo-Re (with) | 17.826 | 0.152 | 3.3 | 14.6 | 3.4 | / |
| Al₂O₃ (with) | 0.707 | 0.578 | 18.6 | 41.4 | 28.8 | 28.7 |
| Glass (with) | 1.148 | 0.483 | 13.9 | 34.3 | 19.9 | 20.8 |
| TiO₂ (with) | 1.522 | 0.345 | 10.4 | 41.3 | 16.1 | 16.8 |
| Mo-Re (with) | 3.741 | 0.225 | 4.8 | 25.6 | 6.0 | 7.0 |
| Al₂O₃ (w/o) | 0.611 | 0.693 | 12.0 | 43.3 | 28.6 | 29.5 |
| Glass (w/o) | 0.672 | 0.706 | 8.1 | 32.5 | 20.6 | 21.4 |
| TiO₂ (w/o) | 1.541 | 0.338 | 9.5 | 40.9 | 15.1 | 16.0 |
| Mo-Re (w/o) | 426 | 2.068 | 2.02 | 4.2 | 2.05 | 2.36 |

$\tau_e^*$: the average lifetime of 2-ExpDec calculated by the weight factor $A_1$ and $A_2$ according to the function:

$$\tau_e = \frac{A_1\tau_1 + A_2\tau_2}{A_1 + A_2}$$

6. The TRPL of dielectric layer passivated samples with and without a post-deposition annealing in air.

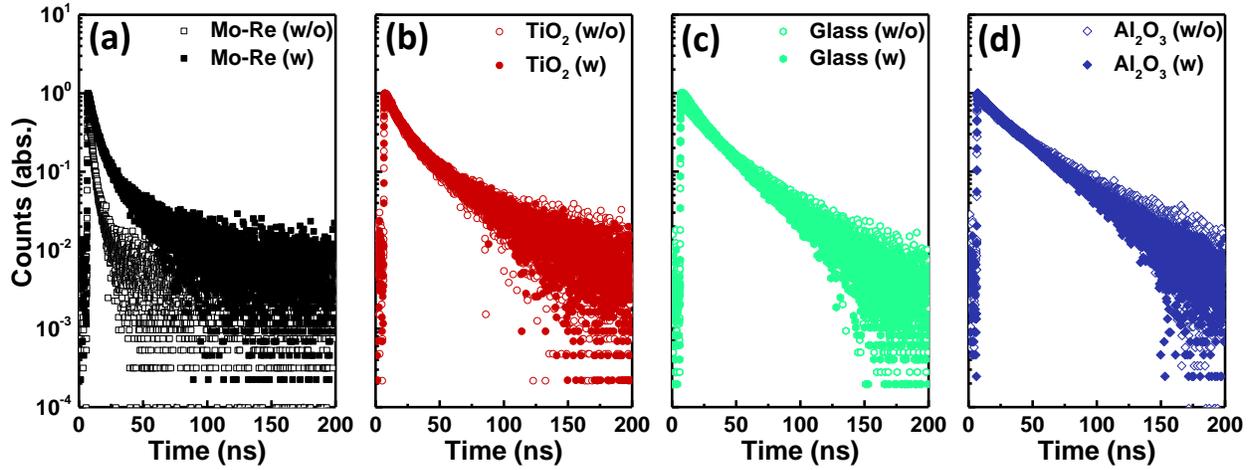

**Figure S4.** Comparison of TRPL measurements of dielectric layer passivated and their unpassivated reference sample with and without a post-deposition annealing in air. Only the unpassivated Mo reference sample has a slower decay and longer life time after the post-deposition annealing. For the passivated samples, they stay almost the same.

7. Optical methods to calculate net doping density of semiconductors.

The method applied to calculate net doping density according to $\Delta E_F$ and minority carrier lifetime.[54] Under dynamic quasi-equilibrium condition, assuming reasonably high mobility, i.e. the photo-generated carriers spread uniformly over the whole film, non-equilibrium electron density $\Delta n$ can be estimated by:

$$\Delta n = G \times \frac{\tau_n^{eff}}{d} \quad (S1)$$

Where $\tau_n^{eff}$ is the electron effective lifetime and $d$ is the absorber thickness, $G$ is the generation flux. The electron density ($n$) and hole density ($p$) are determined by an exponential relation:

$$n = N_c \exp\left(\frac{E_F^n - E_c}{k_B T}\right) \quad (S2)$$

$$p = N_v \exp\left(\frac{E_v - E_F^p}{k_B T}\right) \quad (S3)$$

Where $N_c$ or $N_v$ is the effective density of states of the conduction band or valence band, $E_F^n$ is the quasi-Fermi level of electrons, $E_c$ is the energy of the conduction band minimum, $E_F^p$ is the quasi-Fermi level of holes, $E_v$ is the energy of the valence band maximum and $k_B T$ is the thermal energy. With the bandgap ($E_g$) of the semiconductor we get:

$$E_g = (E_c - E_F^n) + \Delta E_F + (E_F^p - E_v) \quad (S4)$$

For a p-type semiconductor in low excitation, $\Delta n = n$ when $n_0 \ll p_0$ and $p = p_0$, with $p_0$, $n_0$ the equilibrium densities. Combining Equation from S1 to S4, the net doping density (hole density) is determined by:

$$p_0 = \frac{d \cdot N_c N_v}{G \cdot \tau_n^{eff}} \exp\left(\frac{\Delta E_F - E_g}{k_B T}\right) \quad (S5)$$

The values of $N_c$ and $N_v$ we used for this calculation is the same as we applied in simulations that can be found in Table S1.

8. Electron confinement introduced by bandgap gradient.

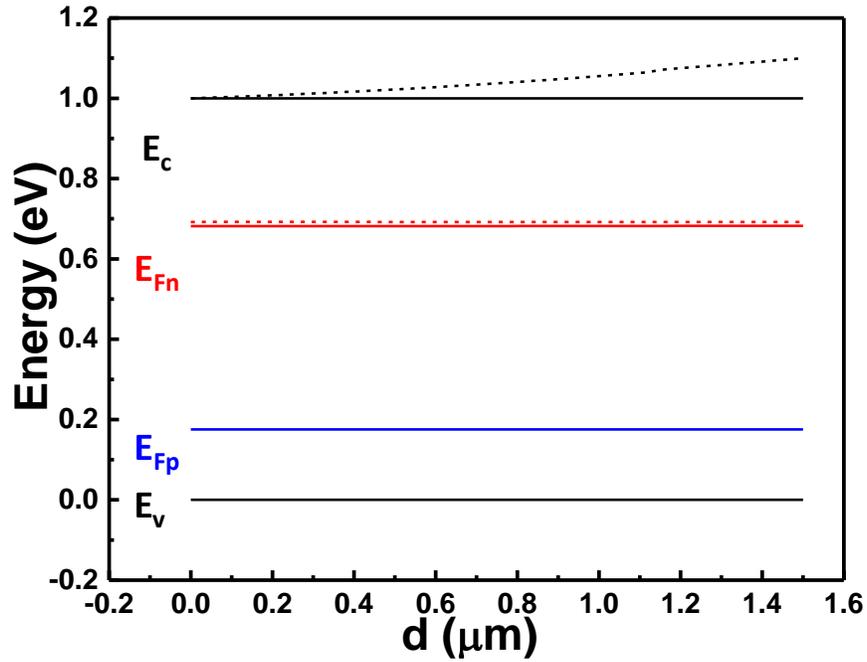

**Figure S5.** The SCAPS simulated band diagram of homogeneous and GBG sample: only a small increase in $\Delta E_F$ ~10 meV is observed, thus we believe the higher $\Delta E_F$ for GBG sample is mainly caused by net doping density rather than conduction band gap gradient introduced electron confinement. With this, we use a homogeneity assumption for the all samples to make the discussion more easily comprehensive. In this simulation, we keep the $S_b = 0$ cm/s (to see the effect of confinement, as opposed to the effect of back surface passivation), $S_f = 1.4 \times 10^3$ cm/s and a constant doping level of $8.8 \times 10^{15}$ cm$^{-3}$.

9. Analytical methods to determine back surface recombination velocity.

For the back and front surface passivated samples, the boundary condition $S_b \gg S_f$ ($S_b \ll S_f$) that leads to Equation 3 is no longer valid. In this case, the general solutions have to be solved by following equations [61, 82]:

$$\frac{1}{\tau_s} = \alpha_0^2 D \tag{S6}$$

Where $\tau_s$ is the surface lifetime due to both surfaces, $D$ is the diffusion constant of minority carriers and $\alpha_0$ is the smallest eigenvalue solution of Equation S7:

$$\tan(\alpha_0 d) = (S_b + S_f)/(\alpha_0 D - \frac{S_b S_f}{\alpha_0 D}) \tag{S7}$$

$S_{b/f}$ is the back or front surface reombination velocity and d is the absorber thinkness. d is the SEM cross-section image determined film thickness of the aborber as shown in Figure S6.

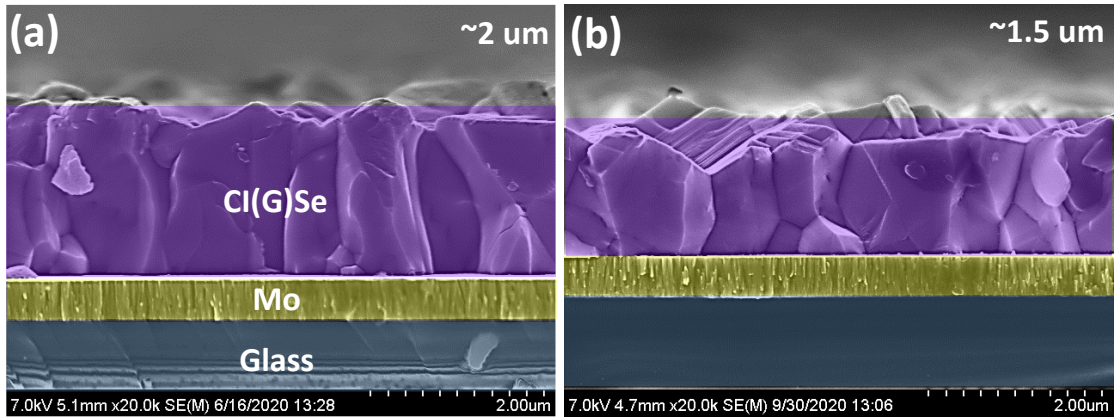

**Figure S6.** The SEM cross section: a) the Mo reference sample of the dielectric layer passivated sample series with a thickness of ~2 μm; b) the no-Ga reference of the GBG sample series with a thickness of ~1.5 μm.

We use the $\tau_{bulk}$ calculated from unpassivated sample with Equation 1 and 3. This $\tau_{bulk}$ should be similar for the passivated samples, because of the same preparation process or recipe of the absorbers. According to the different TRPL measured $\tau_e$, the different $\tau_s$ for each passivated sample can be deduced based on Equation 1. Combining $\tau_s$ of different samples with Equation S6 gives different $\alpha_0$ for each passivated sample. Then, we plot separately the left and right side of Equation

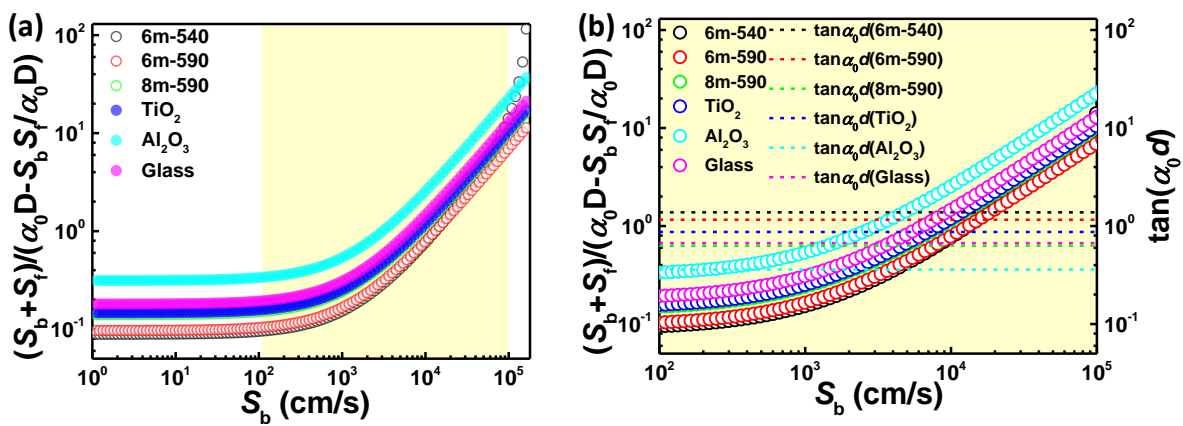

**Figure S7.** The dashed lines shown in b) are the left side term of Equation S7 and circles are the right side term. The $S_b$ can be extracted from the point where they crossover

S7 with respect to $S_b$ in the same figure as shown in **Figure S7**b. The cross point of the two curves gives the $S_b$ value of the passivated sample which makes right side of Equation S7 equal to the left side. The parameters used for calculation and extracted $S_b$ are summarized in Table S4, together with the effective $S_b$ determined from the GGI at the backside (Equation 4 of the manuscript). Considering, that $S_b$ changes over several orders of magnitude, the agreement is very good and confirms the validity of our approach.

**Table S4.** Parameters used for calculation and extracted $S_b$ from an analytical approach

| Sample | d (μm) | $S_f$ (cm/s) | $\tau_b$ (ns) | $\tau_e$ (ns) | $\tan(\alpha_0 d)$ | $S_b$-analytical (cm/s) | $S_b$-Eqn.5 (cm/s) | $S_b$-SAPS (cm/s) |
|---|---|---|---|---|---|---|---|---|
| 6m-540 Ga | 1.55 | 1.4×10³ | 27 | 7.5 | 1.39 | 1.8×10⁴ | 9.3×10³ | 8.2×10³ |
| 6m-590 Ga | 1.55 | 1.4×10³ | 27 | 8.5 | 1.18 | 1.4×10⁴ | 2.1×10⁴ | 1.3×10⁴ |
| 8m-590 Ga | 1.55 | 1.4×10³ | 27 | 13.3 | 0.68 | 4.5×10³ | 6.4×10³ | 3.2×10³ |
| Al₂O₃ | 2.10 | 1.4×10³ | 37 | 28.8 | 0.38 | 3.0×10² | / | 1.2×10³ |
| Glass | 2.10 | 1.4×10³ | 37 | 19.9 | 0.73 | 3.7×10³ | / | 5.9×10³ |
| TiO₂ | 2.10 | 1.4×10³ | 37 | 16.1 | 0.98 | 6.9×10³ | / | 1.3×10⁴ |

## 10. Electrical characterization of GBG samples.

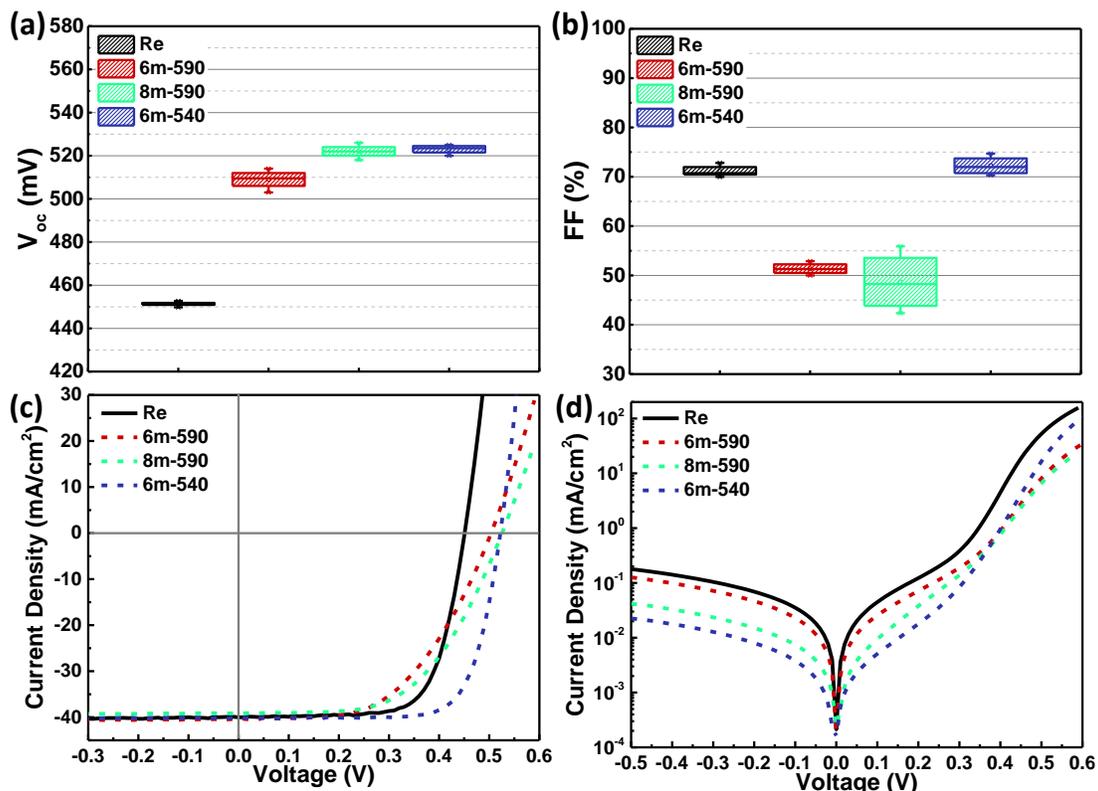

**Figure S8.** The devices' performance of GBG samples. Because without optimizing deposition process of CI(G)Se$_2$, the highest efficiency of 15.7 % is achieved by the low substrate temperature sample (6m-540), even though the high substrate temperature sample (8m-590) has longer minority carrier lifetime and higher $\Delta E_F$. It seems that there is an additional barrier for the transportation of photon generated carriers, which finally introduces extra series resistance that results in an obvious FF loss as shown in figure S8b, c.

Compared to no-Ga sample, most efficiency gain of GBG samples comes from the improvement of $V_{oc}$ as shown in Figure S8a, c. Even though the high substrate temperature sample (8m-590) has a longer minority carrier lifetime and a higher $\Delta E_F$, Figure S8b, c show that the (unoptimized) high temperature process dramatically reduces the fill factor (FF). Irrespective of the FF loss, the lower dark saturation current of GBG samples shown in Figure S8d further proves the good passivation effects of Ga backing grading.

11. Details for a derivation of the ODF.

To make our results and discussions easier to understand, we recapture the crucial derivation from Weiss et al.[7] ODF is directly determined by the slop of the logarithmic Radiation-Generation curve, which is described as:

$$A = \frac{d \ln(R_r)}{d \ln(G)} \tag{S8}$$

Where $A$ is the ODF, $R_r$ is the band to band radiation flux and $G$ is the generation flux, i.e. the illumination intensity. $R_r$ can be described by Planck's generalized law in Boltzmann approximation:

$$R_r \approx \int_0^\infty a(E) \frac{E^2}{4\pi^2 \hbar^3 c^2} \exp\left(-\frac{E}{k_B T}\right) \exp\left(\frac{\Delta E_F}{k_B T}\right) dE = C \exp\left(\frac{\Delta E_F}{k_B T}\right) \tag{S9}$$

Where $a(E)$ is the absorbance, $\hbar$ is the reduced Planck constant, c is the vacuum speed of light, $k_B$ is the Boltzmann constant and T is the temperature of the sample. The constant C represents the integrated part corresponding to the black body radiation. For a p-type semiconductor, in low excitation condition, the electron density equals to the density of non-equilibrium electrons ($n = \Delta n$) which shifts the electron quasi-Fermi level closer to the conduction band with a negligible shift of the hole quasi-Fermi level. This shift of the electron quasi-Fermi level can be calculated by:

$$E_F^n - E_c = k_B T \ln\left(\frac{n}{N_c}\right) \approx k_B T \ln\left(\frac{\Delta n}{N_c}\right) \tag{S10}$$

Where $N_c$ is the effective density of states of the conduction band, $n$ is the total electron density, $\Delta n$ is the non-equilibrium electron density, $E_F^n$ is the quasi-Fermi level of conduction band and $E_c$ is the energy of the conduction band minimum. In a similar way, the hole quasi-Fermi level can be calculated by:

$$E_v - E_F^p = k_B T \ln\left(\frac{p}{N_v}\right) \approx k_B T \ln\left(\frac{N_A}{N_v}\right) \tag{S11}$$

Where $N_v$ is the effective state density of hole, $p$ is the total hole density, $N_A$ is the net doping density, $E_F^p$ is the quasi-Fermi level of valence band and $E_v$ is the energy of the valence band maximum. According to (S1), and (S8) (S11), the ODF in low injection condition can be expressed as:

$$A = \frac{d \ln R_r}{d \ln G} = 1 + \frac{d \ln N_A}{d \ln G} + \frac{d \ln \tau_n^{eff}}{d \ln G} \tag{S12}$$

If there's no additional $N_A$ and $\tau_n^{eff}$ change with generation flux $G$, which results in $A = 1$ in low injection condition.

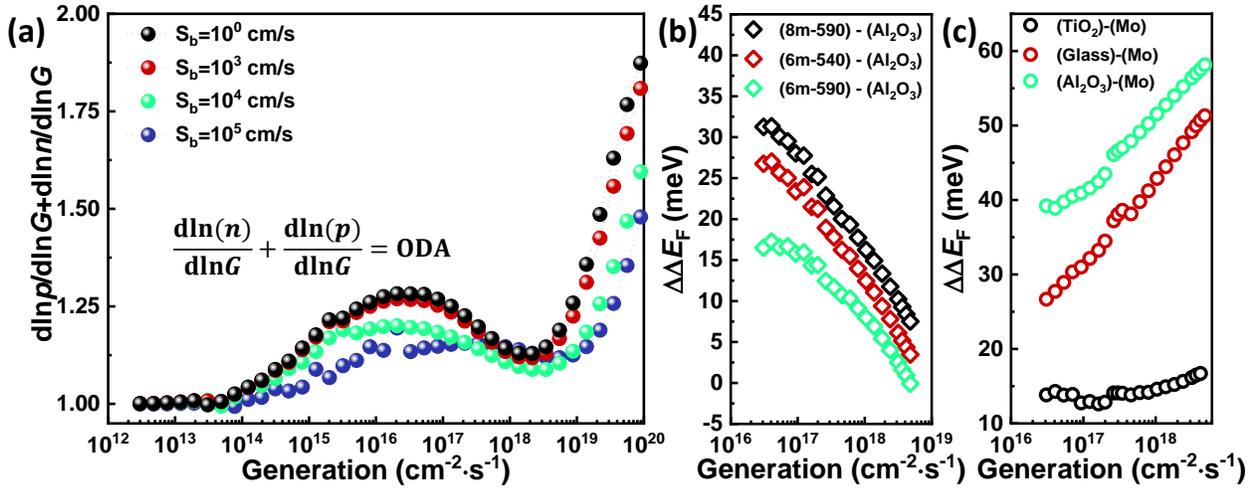

**Figure S9.** a) The back surface recombination can only shift $d \ln p/d \ln G$ to the higher injection level wiout reducing or flatting it, but more electrons recombine with the increase in generation flux which reduces $d \ln n/d \ln G$ (as shown in Figure 7) and consequently the ODF. The higher $S_b$ is, the more electrons recombine with the increase in generation flux and hence causing a smaller ODF; b) c) The $\Delta\Delta E_F$ is experimental $\Delta E_F$ difference between two samples: b) in case of doping level introduced ODF differences, the radiation-generation curves of the samples with different doping level show convergence behavior which also means $\Delta\Delta E_F$ between these samples is getting smaller with increasing generation flux. This is because a higher injection minimizing the doping density differences; c) On the contrary, the back surface recombination dominated samples show a divergence behavior of radiation-generation curves which means $\Delta\Delta E_F$ between these samples is getting larger with increasing generation flux. This is because there are more electrons lossing at a higher injection level through back surface recombination.

12. The ODF derived from different position of the film.

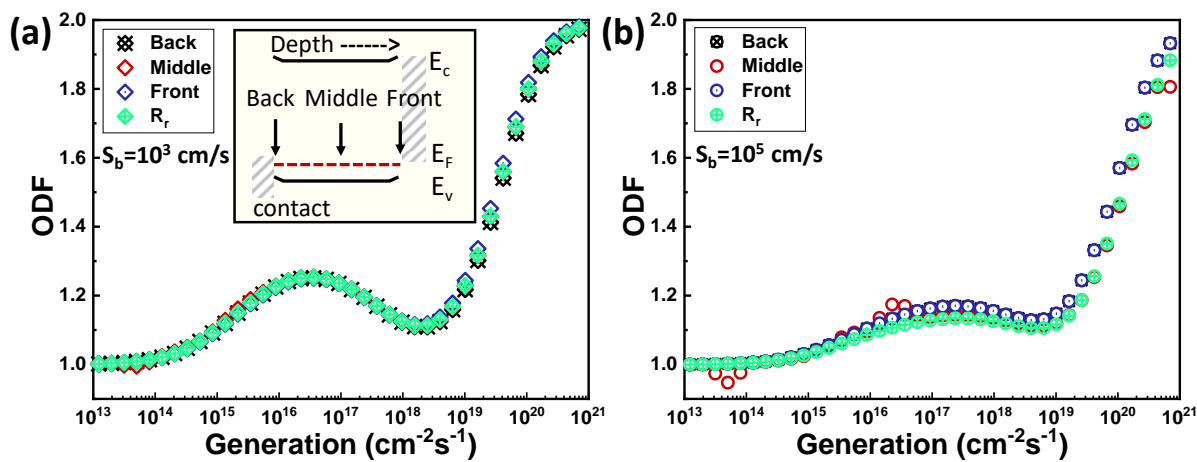

**Figure S10.** The ODF derived from different position of the film based on $\frac{d \ln n}{d \ln G} + \frac{d \ln p}{d \ln G}$ is almost the same to the overall ODF determined by $\frac{d \ln R_r}{d \ln G}$, because the mechanism behind to make the ODF larger than 1 is the same metastable defects transition. It also confirms that using electron and hole density change at certain position of the film, for example at the middle of the film, to represent the ODF of the sample is rational.

13. Electrons occupy the metastable donor states.

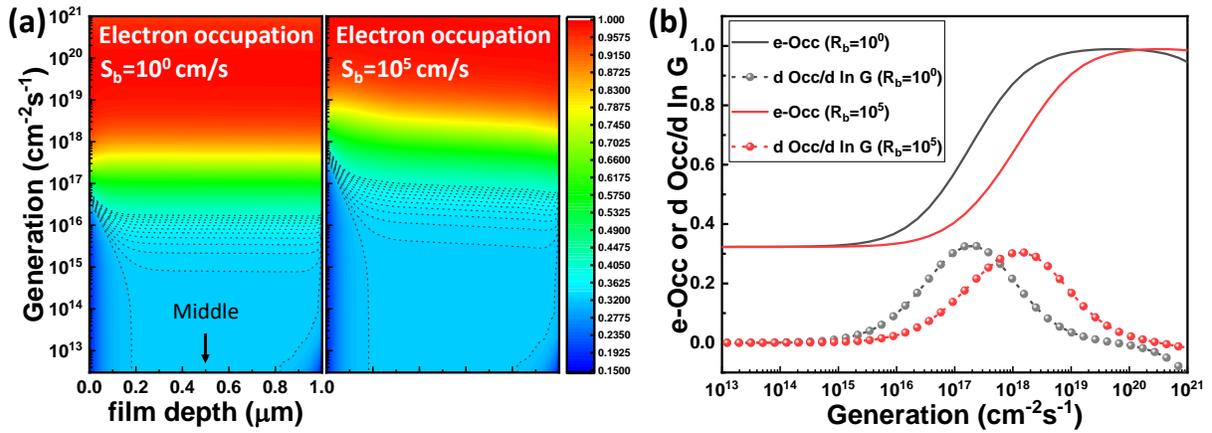

**Figure S11.** a) The occupation of metastable donor states by electrons increases upon generation flux. Compared to a low $S_b$ (left side, $S_b=10^0$ cm/s), a high $S_b$ (right side, $S_b=10^5$ cm/s) shifts the same occupation level to around 1 order of magnitude. The lower occupation at front and back side is due to fixed work function introduced band bending; b) the electron occupation changes with respect to generation flux at middle of the film. The high back surface recombination shift the curve to high generation flux but without reducing its strength.

An electron occupies an metastable donor state means a transformation to acceptor and release a hole ([Metastable donor]$^+$ + e$^-$ → [Metastable acceptor]$^-$ + h$^+$). Compared to the low $S_b$ situation ($S_b=10^0$ cm/s), the high $S_b$ situation ($S_b=10^5$ cm/s) shifts the same level of electron occupation to a higher generation flux around 1 order of magnitude from $10^{17}$ to $10^{18}$ cm$^{-2}$s$^{-1}$. But this shift can't flatten the occupation, which means whenever enough electrons are supplied, the almost same amount of extra hole can be supplied by converting metastable donors to acceptors as shown in **Figure S11**b. The peak of derivative curve in low injection region is shift but the strength of the peak is almost the same. This indicates a shift of $\frac{d \ln p}{d \ln G}$ to the higher generation flux without reducing and flattening it.

## 14. Explanation to illumination intensity dependence of back surface recombination.

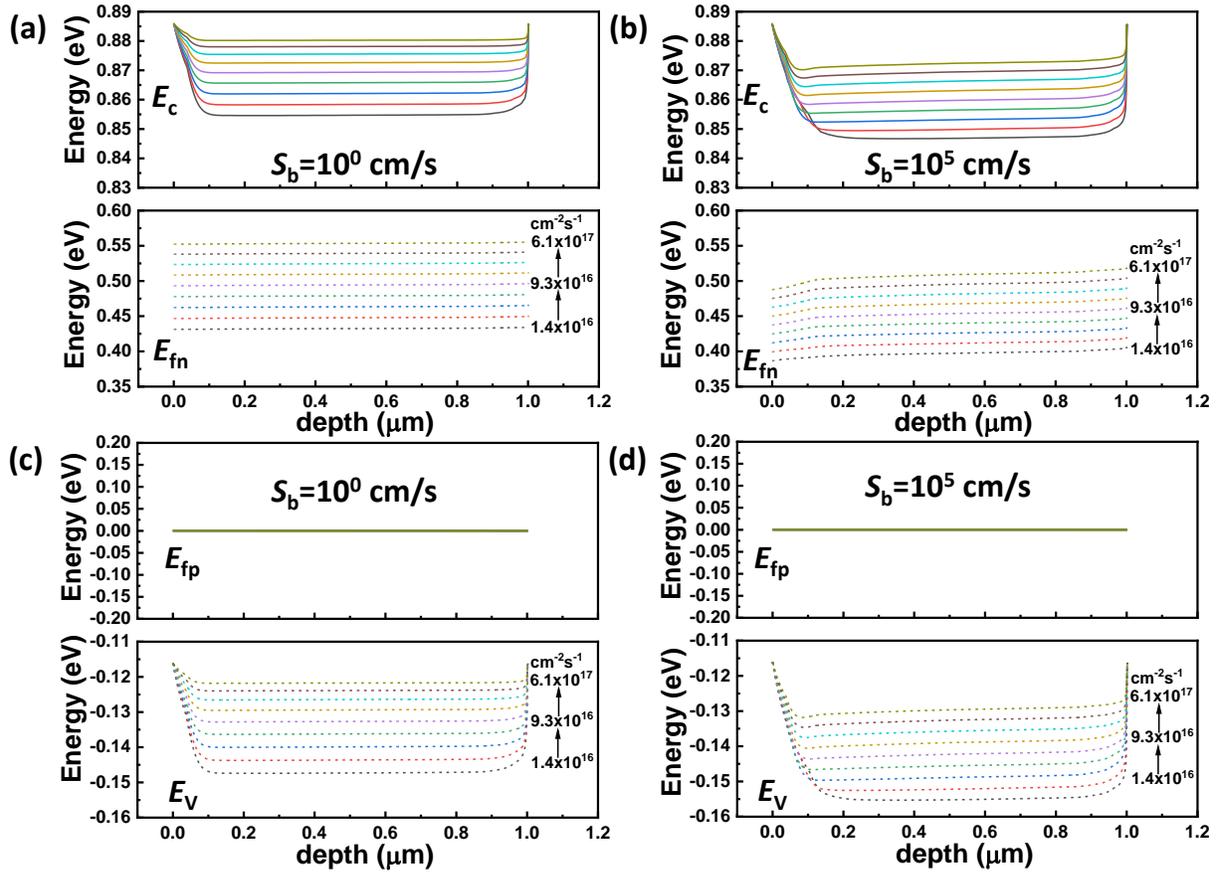

**Figure S12.** Band diagrams of the cases with metastable defects. a) and c) have a low $S_b$ of $10^0$ cm/s, b) and d) have a high $S_b$ of $10^5$ cm/s. The illumination intensity changes from $1.4\times10^{16}$ to $6.1\times10^{17}$ cm$^{-2}$s$^{-1}$. We put the depth dependence of the different energies into separate diagrams to make the small band bendings visible.

In our simulations, a small band bending occurs at the contacts as shown in **Figure S12**a, b. This band bending is caused by the flat-band definition of the back contact in the simulations. The shallow uniform acceptor density ($N_A = 8.8\times10^{15}$ cm$^{-3}$) of defects-free layer is the same as the absorber but without metastable donors to lower the net doping density of the bulk in equilibrium, which means the thin layers' doping density is fixed to $N_A = 8.8\times10^{15}$ cm$^{-3}$. We keep the flat bends setting that makes SCAPS calculate work function of the contacts based on the net doping density of its adjacent layer. As a result, the fixed net doping density of defects free layer fixes the work function. This is a situation similar to the experimental situation, where the Fermi level at the back side is fixed by the work function of the metal back contact of the contacts. But in the bulk, the net doping density ($p_{net}$) is not only determined by $N_A$, but also by the density of metastable donors ($n_{Meta}$) and metastable acceptors ($p_{Meta}$): $p_{net} = N_A + p_{Meta} - n_{Meta}$. In low injection level, metastable donors are still unable to convert to acceptors making $p_{\text{Meta}} < n_{\text{Meta}}$, thus the net doing density of the bulk is lower than $N_A = 8.8 \times 10^{15}$ cm$^{-3}$. It results in both conduction and valence band bending up as shown in Figure S12a,b,c,d, thus having a lower electron density and higher hole density at both contacts compared to the bulk. This situation is like placing a slight p$^+$ layer at both sides. This band bending is flattened with the increase in illumination intensity, because converting metastable donors to acceptors reduces the net doping density difference between bulk and back surface. If we remove the metastable defects but keep the doping difference between bulk and defect free layer in equilibrium, there is almost no change in band bending upon illumination intensity in low injection region as shown in **Figure S13**a,b,c,d. Because in low injection region, the amount of holes gained from generation is too small compared to the doping level. In the case with metastable defects, the hole density at the back surface is controlled by the flattening of $E_v$, which makes the hole density at the back surface ($p_{\text{b-surf}}$) increase much slower when compared to the bulk. As a result, no matter if with or without metastable defects, the $p_{\text{b-surf}}$ stays almost same in low injection region as shown in **Figure S14**a, thus the d ln$p$/d ln$G$ at the back surface stays almost equal to 0 in low injection region as shown in Figure S14b by the black curves. But on the contrary, the amount of extra electrons ($n_{\text{Meta}}^{Ex}$) supplied by the flattening of $E_c$ band bending (Figure S12a,b) due to the metastable defects transition is comparable to free electrons gained from generation, which results in a faster increase in electron density at back surface ($n_{\text{b-surf}}$) in low injection region around $10^{17}$cm$^{-2}$s$^{-1}$ compared to the case without metastable defects configuration as shown in Figure S14a with blue and red curves. It means in low injection region, the $n_{\text{b-surf}}$ of the case without metastable defects configuration follows a linear increase upon illumination intensity, which leads to d ln$n$/d ln$G$ stay almost equal to 1 in low injection region as shown in Figure S14b with red dots, and then a ODF = $\frac{d \ln p}{d \ln G} + \frac{d \ln n}{d \ln G} = 1$ (blue dots in Figure S14b). In case with metastable defects configuration, the $n_{\text{Meta}}^{Ex}$ introduces additional increase in electron density, which makes d ln$n$/d ln$G$ at the back surface larger than 1 in low injection region as shown in Figure S14b with red circle ($S_b = 10^0$ cm/s) and dashed line ($S_b = 10^5$ cm/s), and then a $\text{ODF} = \frac{d\ln p}{d\ln G} + \frac{d\ln n}{d\ln G} > 1$(blue circle and dashed line). Compared to the situation with a low $S_b = 10^0$ cm/s, a $S_b$ of $10^5$ cm/s is high enough to influence $n_{\text{b-surf}}$, as a result it reduces electron density in both cases with and without metastable defects as shown in Figure S14a. But for the case with metastable defects, because of the illumination dependent $n_{\text{b-surf}}$, the back surface recombination turns out to be illumination dependent as well, $[R_{\text{b-surf}}(G) = S_b n_{\text{b-surf}}(G)]$. Consequently, the higher $n_{Meta}^{Ex}$ gain upon illumination intensity introduces more back surface recombination, which flattens the $n_{\text{b-surf}}$ curve and minimizes its difference between the curve without metastable defects as shown in Figure S14a with blue and red dashed lines. Because of this, the d ln$n$/d ln$G$ as well as $\text{ODF} = \frac{d\ln p}{d\ln G} + \frac{d\ln n}{d\ln G}$ curves with respect to illumination intensity is flattened. To offset this additional electron loss at backside due to $n_{Meta}^{Ex}$, more electron in the bulk diffuse to backside causing d ln$n$/d ln$G$ < 1 in the bulk as we mentioned in the text (Figure 9d with solid lines). And we have to emphasize that the existence of metastable defects transition is indispensable for such illumination dependent back surface recombination. Because in both cases (with or without metastable defects configuration), we have almost the same band bending, but the ODF larger than 1 in low injection level only happens to the case with metastable defects, which strongly demonstrates the illumination dependent back surface recombination or the ODF originating from metastable defects transition rather than this artificial band bending.

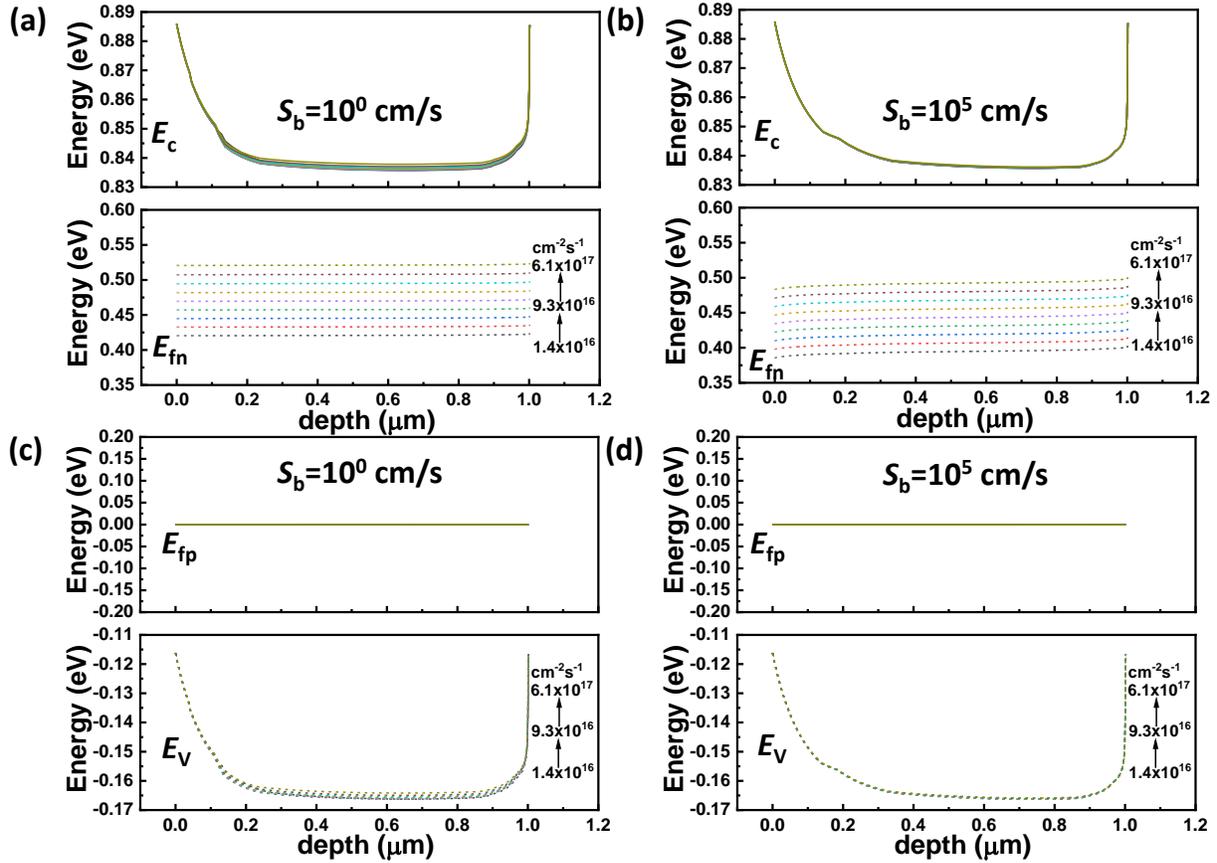

**Figure S13.** Band diagrams of the cases **without** metastable defects. a) and c) have a low $S_b$ of $10^0$ cm/s, b) and d) have a high $S_b$ of $10^5$ cm/s. The illumination intensity changes from $1.4\times10^{16}$ to $6.1\times10^{17}$ cm$^{-2}$s$^{-1}$. These simulations indicate that the change in bandbending and the resulting reduction of the electron concentration, observed in Fig. S12 is due to the presence of metastable defects.

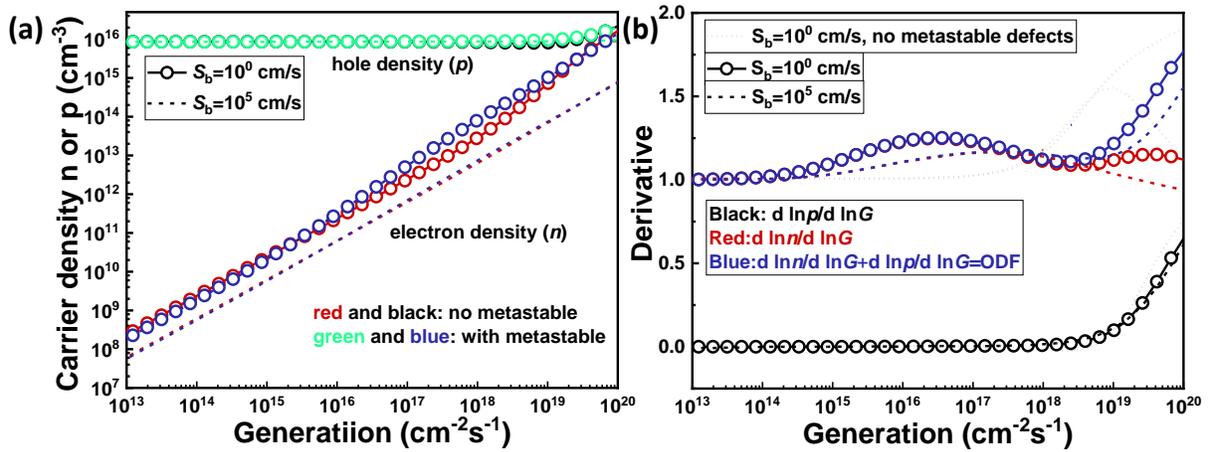

Figure **S14.** Electron and hole density at back surface as a function of generation flux. High back surface recombination velocity, $S_b=10^5$ cm/s, reduces and flattens the extra electron gain at backside due to metastable defects transition in low injection region; b) the derivative of electron and hole density at back surface with respect to generation flux. The high back surface recombination velocity reduces and flattens d ln$n$/ d ln$G$, thus the ODF.

15. Preparing process of absorbers studied in this work

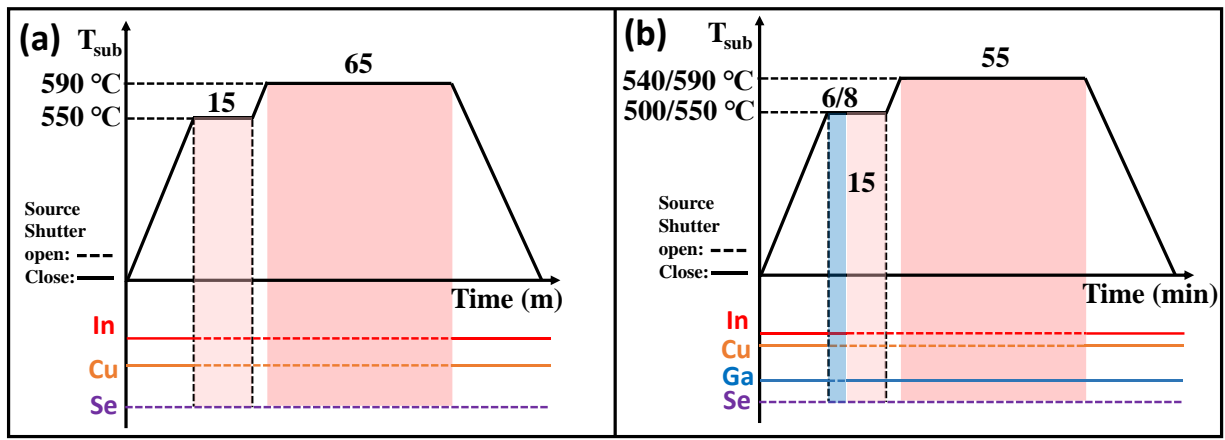

**Figure S15.** a) The absorber preparing process of the dielectric metal oxide layers passivated samples and their reference; b) The absorber preparing process of GBG samples. The Ga distribution is modified by changing the Ga supply duration and substrate temperature. The reference sample is grown with substrate temperature of 550 and 590 °C without Ga supply.